\documentclass[twocolumn]{example2}
\usepackage{bm}

\newcommand{\mytilde}{\raise.19ex\hbox{$\scriptstyle\sim$}}

\received{}
\revised{}
\accepted{}
\submitjournal{}

\graphicspath{{./}{figures/}}

\shorttitle{New Maximum Entropy SL Reconstruction Method}
\shortauthors{Cha \& Jee}

\begin{document}

\title{ MARS: A New Maximum Entropy-Regularized Strong Lensing Mass Reconstruction Method}

\correspondingauthor{M. James Jee}
\email{sang6199@yonsei.ac.kr, mkjee@yonsei.ac.kr}

\author{Sangjun Cha}
\affiliation{Department of Astronomy, Yonsei University, 50 Yonsei-ro, Seoul 03722, Korea}
\author{M. James. Jee}
\affiliation{Department of Astronomy, Yonsei University, 50 Yonsei-ro, Seoul 03722, Korea}
\affiliation{Department of Physics and Astronomy, University of California, Davis, One Shields Avenue, Davis, CA 95616, USA}

\begin{abstract}
Free-form strong-lensing (SL) mass reconstructions typically suffer from overfitting, which manifest itself as false-positive small-scale fluctuations. We present a new free-form MAximum-entropy ReconStruction ({\tt MARS}) method without the assumption that light traces mass (LTM). The {\tt MARS} algorithm enables us to achieve excellent convergence in source positions ($\sim0.001 \arcsec$), minimize spurious small-scale fluctuations, and provide a quasi-unique solution independently of initial conditions.
Our method is tested with the publicly available synthetic SL data {\tt FF-SIMS}. The comparison with the truth shows that  the mass~reconstruction quality is on a par with those of the best-performing LTM methods published in the literature, which have been demonstrated to outperform the existing free-form methods. In terms of the radial mass profile reconstruction, we achieve $<1$\% agreements with the truth for the regions constrained by the multiple images.
Finally, we apply {\tt MARS} to A1689
and find that the cluster mass in the SL regime is dominated by the primary halo centered on the brightest cluster galaxy and the weaker secondary halo also coincident with the bright cluster member
$\mytilde160$~kpc northeast.
Within the SL field, the A1689 radial profile is well-described by a Navarro-Frenk-White (NFW) profile with $c_{200}=5.53\pm0.77$ and $r_s=538^{+90}_{-100}$ kpc and we find no evidence that A1689 is over-concentrated.
\end{abstract}

%%%%%%%%%%%%%%%%%%%%%%%%%%%%%%%%%%%%%%%%%%%%%%%%%%%%%%%%%%%%%%%%%%%%%%%%%%%%%%%%%%%%%%%%%
%%%%%%%%%%%%%%%%%%%%%%%%%%%%%%%%%%%%%%%%%%%%%%%%%%%%%%%%%%%%%%%%%%%%%%%%%%%%%%%%%%%%%%%%%

\section{Introduction} 
\label{sec:intro}
Thanks to the advantage that we do not need any dynamical assumptions, gravitational lensing is now well-accepted as one of the most powerful methods to measure the mass distribution of galaxy clusters \citep[]{2001PhR...340..291B}.
Depending on the characteristic surface mass density of the lensing system, gravitational lensing is broadly classified into two regimes: strong lensing (SL) and weak lensing (WL).

WL uses measurements of coherent shape distortions of background galaxies to infer mass distributions \citep[e.g.,][]{1993ApJ...404..441K,2000ApJ...532...88H, 2006ApJ...642..720J, 2007ApJ...663..717M,
2011ApJ...729..127U, 2017ApJ...851...46F, 2021MNRAS.505.3923S, 2021ApJ...923..101K}. Because of intrinsic shape dispersion of background galaxies, the coherent shape distortions due to lensing are measured through averaging over many source galaxy shapes. Thus, in general the resolution of the WL mass reconstruction depends on the source number density and is limited.

SL occurs in the regime where the surface mass density of the lens exceeds the so-called critical surface mass
density \footnote{Alternatively and perhaps more precisely, one can also define the SL regime as the region where the sum of convergence and shear approaches unity.}, which is defined by the physical surface mass density and the distance ratios among the observer, lens, and source.
SL produces multiple images, which we use for the reconstruction
of the responsible mass distribution.
Although the mass reconstruction is often limited to the central regions of massive clusters, in general one can achieve a much higher resolution and/or tighter constraint than in WL in particular near and inside the multiple image positions. Therefore, SL provides a tremendous opportunity to probe the inner mass profile of the lens in much greater detail \citep[e.g.,][]{2005ApJ...621...53B, 2006ApJ...640..639Z,2007NJPh....9..447J, 2007ApJ...668..643L,2009MNRAS.396.1985Z, 2010ApJ...723.1678C,  2011A&ARv..19...47K,2017ApJ...837...97L,2017MNRAS.472.3177M}.

Since the number of multiple images for a given lens system is finite, inevitably the two-dimensional mass reconstruction is underdetermined. The traditional method resolves the issue by assuming that light traces mass (LTM). The LTM method is implemented by placing some theoretically-motivated analytic halos, such as Navarro-Frenk-White \citep{1996ApJ...462..563N}, Pseudo Isothermal Elliptical Mass Distribution \citep[PIEMD;][]{1993ApJ...417..450K}, etc.,
on the cluster galaxies and adjusting their properties so that the resulting mass map predicts the observed multiple images. Because the properties of these analytic halos are controlled by their parameters, the LTM approach is also referred to as a {\it parametric} mass reconstruction. Typically, the number of free parameters is smaller than the constraints provided by the SL data, and the system is overdetermined. This lack of flexibility often leads to imperfect convergence of multiple image positions in source plane. 

Although the LTM method is a useful approach as a probe of inner mass profiles of galaxy clusters, it is paramount to investigate SL clusters also with the methods that do not rely on the LTM assumption. 
This alternative approach is essential in validating the LTM hypothesis and may be the only route to probe the truth 
if the target cluster indeed deviates from our LTM assumptions.
This non-LTM approach, also referred to as a {\it free-form} or {\it non-parametric} method, typically uses a mesh grid to represent the mass distribution\footnote{Some argue that the term {\it non-parametric} here is misleading because in fact the method requires {\it many} parameters, which
comprise the mesh.}. Compared to the LTM method, the free-form method is flexible because the model is not constrained by shapes and locations of the halos. The flexibility in general allows the method to converge nearly all legitimate multiple images in source plane with negligible scatters.
However, because of the overwhelming number of free parameters compared to
the SL constraints, the system is highly underdetermined, which prevents a unique mass reconstruction and also creates many spurious substructures.

There have been attempts to overcome the drawbacks of the free-form method. One easy solution is to perform the free-form reconstruction numerous times and present the average as the representative mass map.
However, although the average mass map may become much smoother than individual results, one cannot guarantee that the average is a legitimate representative solution. One has yet to prove that the sample distribution (from which the average is derived) is a fair representation of the true parent distribution. Also, the average mass map should be able to predict observed strong lensing data (i.e., ``delens'' multiple image positions in image plane to converge in source plane with negligible scatter and vice versa), which however has not clearly been demonstrated in the literature. 

Another approach to prevent overfitting in the non-LTM method is to employ regularization, which imposes ``Occam's razor" on the final solution. The regularization is often implemented by adding an additional function to the likelihood, which prevents the optimization process from creating spurious small-scale fluctuations unless they are required by the data. This regularization constraint makes the solution unique and smooth. In addition, unlike the {\it average} method above, the solution is legitimate and able to converge multiple image positions in source plane when delensing is performed.

In this study, we propose a free-form MAximum-entropy ReconStruction ({\tt MARS}) method.
The basic maximum-entropy principle is that given a prior, the maximum-entropy distribution
best represents the current state of the information, making the fewest assumptions.
It has been among the most popular regularization schemes
 used for
ill-posed inversion problems and has been applied to WL and WL+SL mass reconstruction studies \citep[][]{1998MNRAS.299..895B, 1998A&A...337..325S, 2007ApJ...661..728J}.
The current approach in this paper is similar to the WL+SL study of \citet[][hereafter J07]{2007ApJ...661..728J} with some important updates. First,
our grid represents the mass distribution of the SL system, not the lensing potential. 
Our experiment shows that the difference makes the optimization of the target function (likelihood+regularization) much more stable than the case where the lensing potential is used. Since the relation between the mass and deflection field is non-local, we have to extend our reconstruction field slightly larger than the target reconstruction field in order to account for the contribution from the mass outside the target field.
Second, we update the prior normalization scheme used for the cross-entropy computation. This reduces the artifact that flattens the mass peak more than we desire.
Third, we employ a new optimization engine, which functions reasonably well even in the case where the number of free parameters is huge ( $\mytilde19,600$).

This paper is structured as follows. In \textsection\ref{sec:method}, we describe our algorithm for the SL mass reconstruction and error estimation. In \textsection\ref{sec:data}, we introduce the synthetic SL data that we use to evaluate the performance of our algorithm and real observational SL data of A1689.
We demonstrate that the algorithm produces high-fidelity mass reconstruction results for the synthetic clusters and also discuss the result for A1689 in \textsection\ref{sec:result}. \textsection\ref{sec:discussion} discusses the dependence of the mass reconstruction on the grid resolution and initial condition. We conclude in \textsection\ref{sec:conclusion}. Unless stated otherwise, this paper assumes the cosmology with $\Omega_{m}=0.27$, $\Omega_{\Lambda}=0.73$ and $h=0.72$.

%%%%%%%%%%%%%%%%%%%%%%%%%%%%%%%%%%%%%%%%%%%%%%%%%%%%%%%%%%%%%%%%%%%%%%%%%%%%%%%%%%%%%%%%%
%%%%%%%%%%%%%%%%%%%%%%%%%%%%%%%%%%%%%%%%%%%%%%%%%%%%%%%%%%%%%%%%%%%%%%%%%%%%%%%%%%%%%%%%%

\section{Method} \label{sec:method}
\subsection{{\tt MARS} Algorithm}\label{subsec:reconstruction_method}
{\tt MARS} is based on a free-form source plane minimization with maximum entropy regularization.
Here, we briefly summarize the algorithm unique to our method. For general details of the SL mass reconstruction, readers are referred to excellent review papers \citep[e.g.,][]{kochanek2006, 2011A&ARv..19...47K}. 

Gravitational lensing is a nonlinear mapping that transforms the source position $\bm{\beta}$ to the image position $\bm{\theta}$ via the following lens equation:
\begin{equation}
    \bm{\beta}=\bm{\theta}-\bm{\alpha}(\bm{\theta}),
\label{eq_alpha}
\end{equation}
where $\bm{\alpha}$ is the deflection angle.
One can compute the deflection angle \bm{$\alpha$} via either the derivative of the deflection potential $\Psi$:
\begin{equation}
\bm{\alpha}=\nabla \Psi 
\label{eqn_deflection_via_der}
\end{equation} 
or the convolution of the convergence $\kappa$:
\begin{equation}
    \bm{\alpha} (\bm{\theta}) = \frac{1}{\pi} \int
    \kappa (\bm{\theta}^{\prime}) \frac{\bm{\theta}-\bm{\theta}^{\prime}}{|\bm{\theta}-\bm{\theta}^{\prime}|^{2}} \bm{d^{2} {\theta}}^{\prime}. 
    \label{eqn_deflection_via_con}
\end{equation}
The convergence $\kappa$ is the dimensionless surface mass density defined as:
\begin{equation}
    \kappa=\frac{\Sigma}{\Sigma_{c}},
\label{eqn_kappa}
\end{equation}
where $\Sigma$ ($\Sigma_c$) is the (critical) surface mass density.
The $\Sigma_c$ is defined as:
\begin{equation}
    \Sigma_{c}=\frac{{c^2}D_s}{4{\pi}G{D_{ds}}{D_d}},
\end{equation}
where $D_{ds}$ is the angular diameter distance between the source and lens, $D_s$ is to the source, and $D_d$ is to the lens.

The first method (obtaining the deflection angle $\bm{\alpha}$ by taking the derivative of the lensing potential $\Psi$, eqn~\ref{eqn_deflection_via_der}) has been advocated over the second one
(eqn.~\ref{eqn_deflection_via_con}) 
by some authors \citep[e.g.,][]{1998A&A...337..325S, 2005A&A...437...39B, 2007ApJ...661..728J}, who claim that the second method leads
to biased results because the mass distribution outside the reconstruction field can affect the evaluation of the deflection angle (eqn.~\ref{eqn_deflection_via_con}).
However, this weakness of the second method can be overcome by simply and sufficiently extending the reconstruction field (as we demonstrate in \textsection\ref{subsec:result_mock_clusters}).
In this paper, we choose the second method because our experiment shows that the function minimization is much more stable than when we use the first method, although the field extension increases the number of free parameters by a significant factor.

Both $\kappa$ and $\bm{\alpha}$ are functions of the source redshift.
Since our sources are at different redshifts, we first evaluate them  at a reference redshift $z_f$ and
then scale the deflection angle for each source according to its redshift as follows: 
\begin{equation}
    \bm{\alpha}(z_s)=\bm{\alpha}(z_{f})W(z_{f},z_s),
\end{equation}
where $W(z_{f}, z_s)$ is the cosmological weight:
\begin{equation}
    W(z_{f},z_s)=\frac{D(z_{f})D(z_{l},z_s)}{D(z_s)D(z_{f},z_{l})}.
\end{equation}
$D(z)$ is the angular diameter distance to the redshift $z$ while $D(z_A, z_B)$ is between the redshift $z_A$ and $z_B$. In this paper, we set $z_f=9$.

All multiple images from each source should converge to the same positions in the source plane when "delensed" and we minimize
their scatters:
\begin{equation}
    \chi^{2}=\sum_{i=1}^{I} \sum_{j=1}^{J}\sum_{k=1}^{K}\frac{(\bm{\theta}_{i,j,k}-\bm{\alpha}_{i,j,k}(z)-\bm{\beta}_{k})^{2}}{{\sigma_{i}}^{2}},
\label{eqn_chi_squared}
\end{equation}
where
\begin{equation}
    \bm{\beta}_{k}=\frac{1}{J}\sum_{j=1}^{J}(\bm{\theta}_{i,j,k}-\bm{\alpha}_{i,j,k}(z))
\end{equation}
and
$I$ is the total number of sources, $J$ is the total number of multiple images for each source, and $K$ is the total number of ``knots" for each multiple image.
Here, ``knots" refer to the positions of distinctive morphological features within a source if identifiable, which also must converge if the mass reconstruction is flawless. In this paper, we do not utilize the information 1) because our synthetic SL
data do not provide them and 2) because A1689 have only a few sources with clear knots. Nevertheless, as explained below, we set up {\it virtual} knots as a convenient way to evaluate local magnifications.
The notation $\sigma_{i}$ in eqn.~\ref{eqn_chi_squared} is the maximum scatter of the multiple image positions from the $i^{th}$ source in the source plane and is given by:
\begin{figure}
\epsscale{0.75}
    \plotone{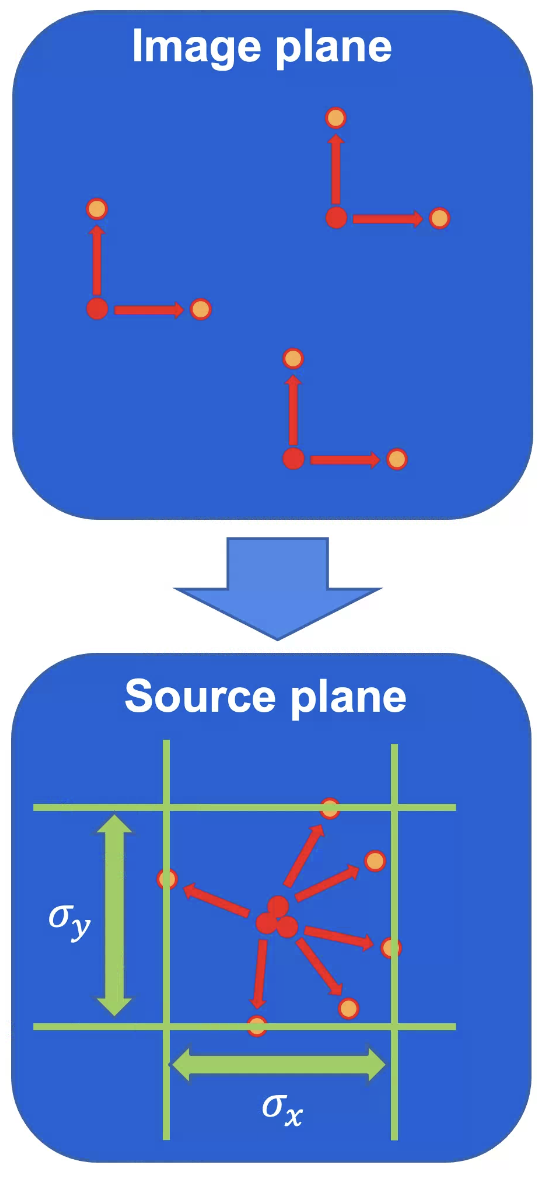}
    \caption{Virtual knot and measurement of maximum scatter.
    Red dots (Orange dots) indicate the fiducial centers (virtual knots) of the multiple images from a source.
    We use the maximum scatter of the virtual knot positions $\bm{\beta}$ in the source plane as an input to the $\chi^2$ term. Since the maximum scatter is not fixed, but updated per $\chi^2$ evaluation and this prevents the model from favoring a high magnification solution.}
    
    \label{fig:sigma}
\end{figure}

\begin{equation}
    \sigma_{i}=\sqrt{\sigma_{x,i}^2 + \sigma_{y,i}^2},
\label{eqn_sigma}
\end{equation}
where $\sigma_{x,i}$ and $\sigma_{y,i}$ are the largest differences in $\bm{\beta}$ along the $x$- and $y$-axes, respectively.
If we had used a fixed value for $\sigma_i$ (for example, $\sigma_i=0.001\arcsec$), the mass reconstruction would have been biased toward high-magnification results because a higher magnification model can  make the $\chi^2$ term smaller.
Using the current scheme prevents the converging model from favoring a high magnification solution.
Although one can consider a full lensing Jacobian matrix and compute the local magnification with it, in this paper, we implement this by simply creating two virtual knots $0\farcs2$ from the source center in such a way that the two vectors connecting the knots and the center are perpendicular to each other.
Figure~\ref{fig:sigma} illustrates the concept of our virtual knots in the image plane and the way that we evaluate $\sigma_{x,i}$ and $\sigma_{y,i}$ in the source plane.

The following cross-entropy\footnote{This term is also referred to as the Kullback-Leibler divergence \citep{KL1951}.}
is used to regularize our mass reconstruction to achieve smooth results and prevent overfitting:
\begin{equation}
    R=\sum_{m,n}\left (p(m,n)-\kappa(m,n)+\kappa(m,n)\mathrm{ln}\frac{\kappa(m,n)}{p(m,n)} \right),
\label{cross_entropy}
\end{equation}
where $p(m,n)$ and $\kappa(m,n)$ are the prior and convergence at the position $(m,n)$, respectively.
This regularization is different from the one in \citet{2007ApJ...661..728J}, where the the first two terms in Equation~\ref{cross_entropy} are dropped and $p$ and $\kappa$ are normalized in such a way that the summation of each becomes unity. 
We find that this modification reduces the artificial flattening of the mass peak.

Our final target function to minimize is then:
\begin{equation}
    f={\chi}^{2}+rR, \label{eqn_f}
\label{total_equation}
\end{equation}
where $r$ is a regularization control parameter to adjust the balance between ${\chi}^{2}$ and $R$.
Choosing too large values leads to oversmoothed results whereas too small values cause overfitting. In the latter case, the resulting mass map is not uniquely determined and shows a number of (unphysical) small-scale features. Since equation~\ref{eqn_f}, composed of the log-likelihood and prior terms, is based on the Bayesian principle, one may suggest a value, which makes the $\chi^2$ value per component approach unity (e.g., Seitz et al. 1998). However, in our case, since $\sigma_i$ (eqn.~\ref{eqn_sigma})  in the $\chi^2$ term (eqn.~\ref{eqn_chi_squared}) is not a $\mytilde68$\% error estimate, but a maximum dispersion, this scheme is not directly applicable. Instead, we examine the relation between $r$ and the resulting mean scatter in the source plane and choose the value in such a way that the mean scatter becomes $\mytilde0.001\arcsec$. We discuss the mass map dependence on the $r$ parameter in \textsection\ref{subsec:discussion_r}.

Our mass reconstruction is performed in two steps. In the first one, we embed a $50\times50$ target reconstruction field in the center of a $70\times70$ grid, where the 10 pixel margin surrounding the target field is introduced to include the impact of the mass outside the $50\times50$ target reconstruction field.
After the minimization of the function $f$ in the first step, 
we double the resolution ($140\times140$) and refine the result from the initial $70\times70$ solution. As explained below, our mass reconstruction is computationally intensive. This two-step implementation is found to speed up the convergence substantially compared to the case
where we start with the $140\times140$ grid from the beginning.
We considered different sizes for the margin and found that larger margins do not change the final result significantly.
Because the observed multiple images do not lie exactly on the lattice point of the grid, we use bi-cubic interpolation to compute $\bm{\alpha}$ at the image positions.

We start the function minimization from a flat prior. If the minimization is finished for the given prior, we restart the minimization after updating the prior with the new convergence map obtained in the previous minimization.
This prior needs to be slightly smoothed to prevent overfitting. 
We applied a Gaussian convolution kernel whose smoothing scale is \bm{$\sigma=0.6$} (1.2) grid cell for the \bm{$70\times70$} (\bm{$140\times140$}) grid.
This minimization cycle is repeated until there is no improvement. 

Since the number of free parameters is extremely large ($4,900$ and $19,600$ for the 70$\times$70 and 140$\times$140 grids, respectively) and the function is highly nonlinear, caution is needed when we select a minimization algorithm.
We choose the L-BFGS-B algorithm \citep{Byrd1995, 10.1145/279232.279236}, which is based on the quasi-Newton method. The algorithm does not require a full Hessian matrix to determine descent directions and this is a significant merit in our study because the full Hessian matrix evaluation is prohibitively expensive.

%%%%%%%%%%%%%%%%%%%%%%%%%%%%%%%%%%%%%%%%%%%%%%%%%%%%%%%%%%%%%%%%%%%%%%%%%%%%%%%%%%%%%%%%%

\subsection{Uncertainties of Mass Maps}\label{subsec:uncertainties}
It is important to understand the limitation of the SL mass reconstruction in order to interpret the result quantitatively and compare it with other studies. In the LTM approach, it is straightforward to compute the parameter uncertainties because the number of free parameters is small and the system is in general over-determined.
However, since the quoted uncertainties are valid only within the LTM assumptions, one must include systematic errors due to the chosen LTM models when assessing the validity of the result.

In the non-LTM or free-form method, computation of the parameter uncertainties is non-trivial. In particular, if the minimization function does not contain any regularization term, the system is highly under-determined and thus the solution is not unique. 
To bypass the difficulty, some authors perform many minimization runs and quote the standard deviation of these solutions. However, since the mean is not one of the valid solutions, it is questionable whether  
the standard deviation can be used as a reliable uncertainty proxy.

Our {\tt MARS} free-form method provides a unique solution because the maximum-entropy term makes the system over-determined. Obviously, this is a significant advantage over other free-from methods and allows us to estimate the parameter (i.e., mass cell) uncertainties from the Hessian matrix. Although the matrix is extremely large ($70^2\times70^2$ and $140^2\times140^2$ for the 70$\times$70 and 140$\times$140 grids, respectively), the computation is needed only once after the convergence is reached. Also, since evaluation of each matrix element is independent, computation of the entire matrix elements can be easily parallelized.
This Hessian-based uncertainty estimation is also used in \citet{2007ApJ...661..728J}. However, their parameters are potentials and thus complex error propagation using all covariances
was needed to obtain mass uncertainties.  
Our Hessian matrix is a square matrix whose elements are the second derivatives\footnote{We computed the second derivatives numerically with respect to each grid cell.} of the function $f$ (eqn.~\ref{total_equation}).
Under the assumption that the posterior of each $\kappa$ value follows the Gaussian distribution, we can use the inverse of the Hessian matrix to
obtain the covariance of $\kappa$. 

The covariance matrix needs to be re-scaled if
the $\chi^2$ value (\ref{eqn_chi_squared})
per multiple image departs from unity when the convergence is reached. We adopted $\sigma=0\farcs02$ to be our fiducial mean source plane scatter and multiplied the per-image $\chi^2$ value to our covariance matrix. In \textsection\ref{subsec:unc_change}, we discuss how the uncertainty depends on our choice for the mean source plane scatter.

We note that the uncertainty map computed in this way has some limitations. First, it is derived under the assumption that the error distribution is Gaussian. Also, the values are valid only for the given prior. Finally, our uncertainty map does not include systematic errors due to the regularization and other method-specific artifacts.

%%%%%%%%%%%%%%%%%%%%%%%%%%%%%%%%%%%%%%%%%%%%%%%%%%%%%%%%%%%%%%%%%%%%%%%%%%%%%%%%%%%%%%%%%
%%%%%%%%%%%%%%%%%%%%%%%%%%%%%%%%%%%%%%%%%%%%%%%%%%%%%%%%%%%%%%%%%%%%%%%%%%%%%%%%%%%%%%%%%

\section{Data} \label{sec:data}
We apply {\tt MARS} to three galaxy clusters.
Two of them are the mock clusters from \citet[][hereafter M17]{2017MNRAS.472.3177M} and the other is A1689. Testing our algorithm with the mock clusters are important because this allows us to objectively evaluate the performance by comparing the reconstruction with the ``truth." Also, several existing algorithms have participated in the mass reconstruction of the same mock clusters and this enables straightforward comparisons. 

Most of the results presented in M17 were submitted before unblinding. Therefore, one may argue that any results published after unblinding should not be evaluated on equal footing. For LTM-based methods, which can utilize the true halo locations, shapes, slopes, etc. as their initial conditions or constraints, the solutions can be optimized or improved after the truth is unblinded. However, when it comes to grid-based free-form methods like MARS, these “knobs” do not exist and thus in general it is difficult to utilize the unblinded data. For certain free-form methods that produce different results for different runs/realizations, confirmation bias can occur because one can subjectively discard the samples that are distant from the unblinded truth. Since our mass reconstruction produces a quasi-unique solution regardless of the initial condition/assumption, this confirmation bias is irrelevant to our algorithm.

%%%%%%%%%%%%%%%%%%%%%%%%%%%%%%%%%%%%%%%%%%%%%%%%%%%%%%%%%%%%%%%%%%%%%%%%%%%%%%%%%%%%%%%%%

\subsection{Mock Cluster Data} \label{subsec:mock_clusters}
\begin{figure*}
\centering
\includegraphics[width=0.8\textwidth]{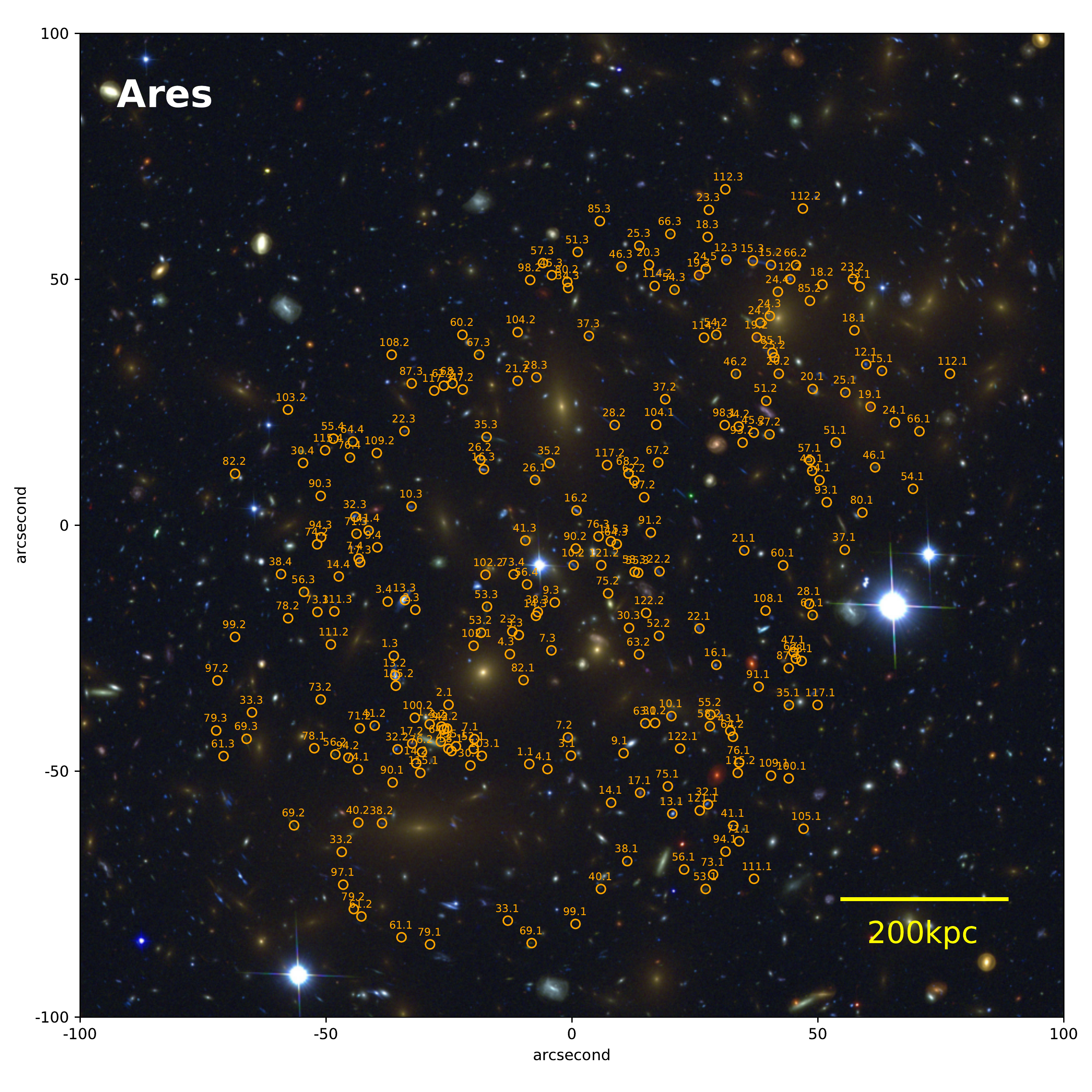}
\caption{Multiple image distribution of the mock galaxy cluster Ares. The $200\arcsec \times 200\arcsec$ color composite image
is created with the mock ACS data with F435W, F606W, and F814W filters. Circles and numbers represent the locations and IDs of the multiple images. We used same coordinate system as M17, up (right) is north (west).}
\label{fig:ares_all}
\end{figure*}

\begin{figure*}
\centering
\includegraphics[width=0.8\textwidth]{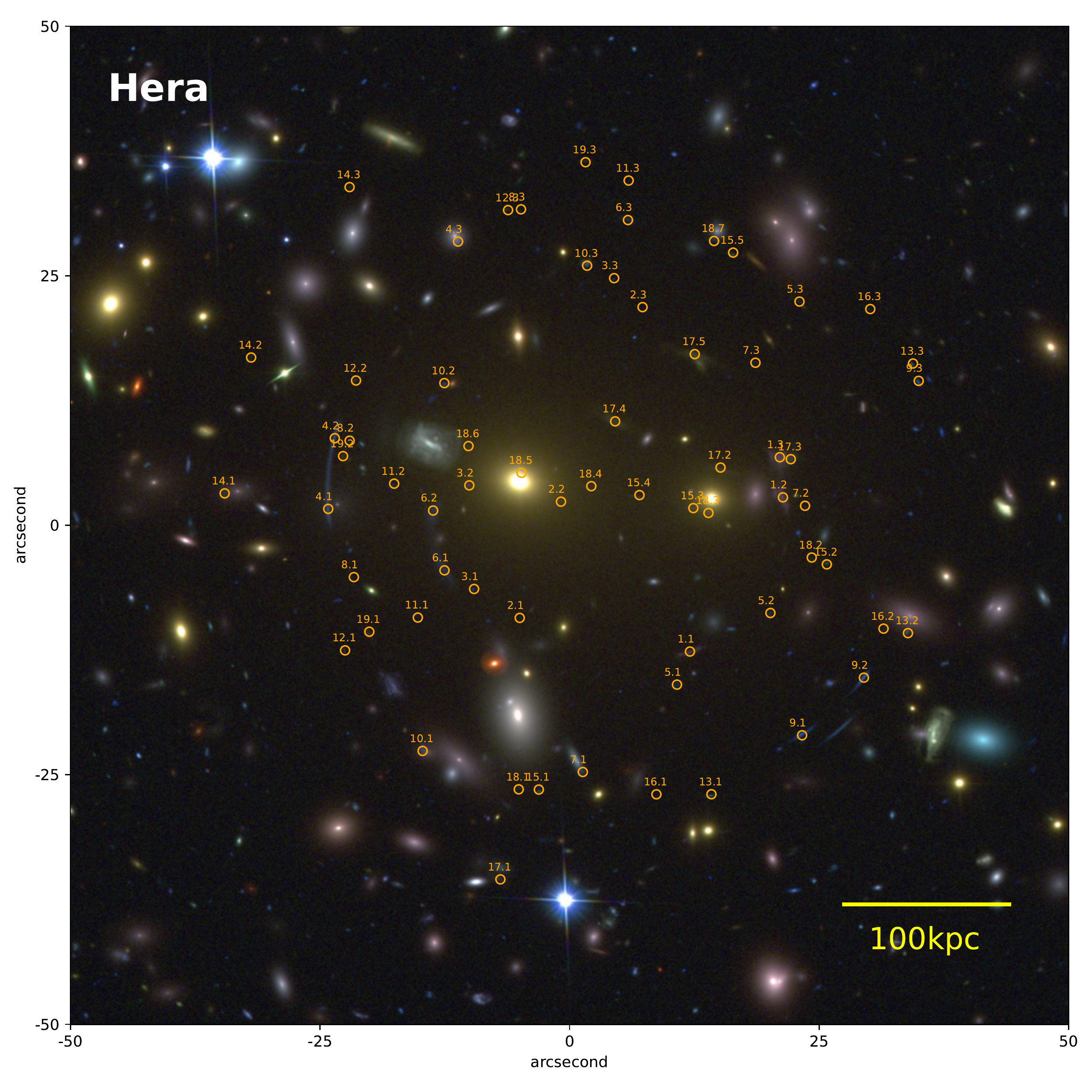}
\caption{Multiple image distribution of the mock galaxy cluster Hera. Same as Figure~\ref{fig:ares_all} except that the field of view is $100\arcsec \times 100\arcsec$.}
\label{fig:hera_all}
\end{figure*}

To test our SL reconstruction method, we use the two mock clusters from M17, whose SL data are publicly available.\footnote{\url{http://pico.oabo.inaf.it/~massimo/Public/FF/index.html}} 
During the review process of the current article, we were informed that the ``true" deflection field data released by M17 are not self-consistent with their multiple image catalogs. Our independent investigation reveals that indeed   the scatters in the source plane measured with
their ``true" deflection field data are as large as 
$\mytilde0.1\arcsec$ for both clusters. Note that the results in this paper are obtained with the same multiple image catalogs available to the M17 participants.
Here we provide only a brief summary on how these two mock clusters are generated. For detail, we refer readers to M17.

The mock galaxy cluster $Ares$ at $z=0.5$ is created with the semi-analytic code {\tt MOKA} \citep{2012MNRAS.421.3343G}. $Ares$ consists of
three components: 1) the main dark matter halo, 2) cluster members, and 3) the brightest cluster galaxy (BCG), which all follow analytic profiles.
The SL data are produced under the flat $\Lambda$CDM model with 
$\Omega_{m}=0.272$ and $H_{0}=70.4~ \mbox{km}~\mbox{s}^{-1} \mbox{Mpc}^{-1}$. The projected mass distribution shows many sharp ``spikes" at the location of galaxies due to the cuspy centers of the analytic profile halos.

The other mock galaxy cluster $Hera$ at $z=0.507$ is generated from a high-resolution  ($m_{DM}=10^{8}h^{-1} M_{\sun}$) zoom-in 
re-simulation of the initial low-resolution cosmological simulation of \cite{2014MNRAS.438..195P}, which  assumes
a flat $\Lambda$CDM with $\Omega_{m}=0.24$, $\Omega_{b}=0.04$, $\sigma_8=0.8$, and $H_{0}=72~ \mbox{km}~\mbox{s}^{-1} \mbox{Mpc}^{-1}$.  Compared to $Ares$, the galaxy-scale structures
are somewhat smoothed because of the finite resolution of the simulation.

$Ares$ and $Hera$ differ greatly in mass. $Ares$ is comprised of two halos, whose virial masses
are $1.32\times10^{15} h^{-1} M_{\sun}$ 
and $8.8\times10^{14} h^{-1} M_{\sun}$
for the southeastern and northwestern halos, respectively.
The projected distance between the two halos is $\mytilde400~h^{-1}$~kpc.
On the other hand, the total mass of $Hera$ is $9.4\times10^{14} h^{-1} M_{\sun}$.
This cluster is also bimodal as in the case of $Ares$, comprised of the eastern and western halos.
The projected separation is much smaller ($\mytilde130~h^{-1}$~kpc) than $Ares$.

Since $Ares$ is more massive than $Hera$, the 
angular extent of the multiple image distribution is also larger.
For $Ares$, we choose the central $200\arcsec \times 200\arcsec$ region as our reconstruction field, which contains 242 multiple images from 85 sources. For $Hera$, we reconstruct the central $100\arcsec \times 100\arcsec$ region and there are 65 multiple images from 19 source galaxies. 
We display the color composite images of $Ares$ ($Hera$) and its multiple image distribution in Figure~\ref{fig:ares_all} (\ref{fig:hera_all}).

%%%%%%%%%%%%%%%%%%%%%%%%%%%%%%%%%%%%%%%%%%%%%%%%%%%%%%%%%%%%%%%%%%%%%%%%%%%%%%%%%%%%%%%%%

\subsection{Abell 1689 Data} 
\label{subsec:real_cluster}
\begin{figure*}
\centering
\includegraphics[width=0.9\textwidth]{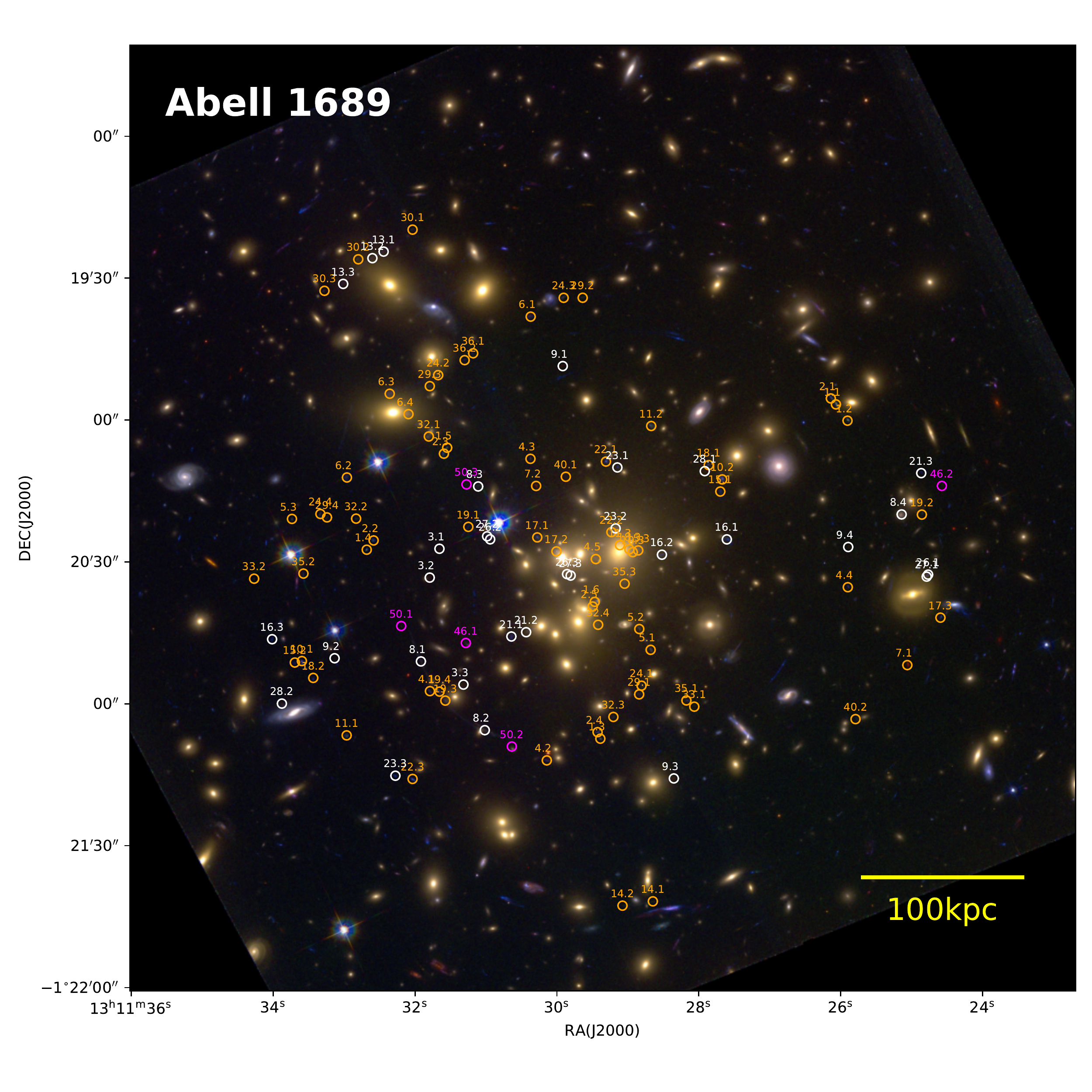}
\caption{Multiple images distribution of A1689.
We display the central $200\arcsec \times 200\arcsec$ region, which is our mass reconstruction field. Orange (white) circles and numbers
indicate the positions and IDs of ``gold" (``silver") images. Magenta is used to annotate ``gold candidate". 
The color-composite image is created with the ACS F475W, F625W, and F775W filters.}
\label{fig:A1689_all}
\end{figure*}

The galaxy cluster A1689 at $z=0.18$ has been a subject of a number of SL studies \citep[e.g.,][]{2005ApJ...621...53B,2006ApJ...640..639Z, 2007ApJ...668..643L,2010ApJ...723.1678C,2015MNRAS.446..683D}.
We compiled the multiple images from the following three studies: \citet[][hereafter B05]{2005ApJ...621...53B}, \citet[][hereafter L07]{2007ApJ...668..643L},
and \citet[][hereafter C10]{2010ApJ...723.1678C}.
The multiple images of A1689 used in the current study is listed in Table~\ref{table1}, which
contain 109 multiple images from 34 source galaxies.
We classify them into the ``gold", ``silver", and ``gold candidate" samples.
The gold images have spectroscopic redshifts. The silver images have only photometric redshifts, which are agreed to be bona fide multiple images
by all three papers.
The gold candidates are the multiple images whose spectroscopic redshifts were unknown to the authors in the three studies, but later measured by \citet[][hereafter B16]{2016A&A...590A..14B}.
There are 73 multiple images from 22 source galaxies in the gold sample, 31 multiple images from 10 source galaxies in the silver sample, and 5 multiple images from 2 source galaxies in the gold candidate sample.
We display the multiple image distribution used in this study in Figure ~\ref{fig:A1689_all}, which shows the central $200\arcsec\times200\arcsec$ region, where we perform the SL mass reconstruction.
We follow the numbering scheme of B05 while adopting the coordinates of C10.

%%%%%%%%%%%%%%%%%%%%%%%%%%%%%%%%%%%%%%%%%%%%%%%%%%%%%%%%%%%%%%%%%%%%%%%%%%%%%%%%%%%%%%%%%
%%%%%%%%%%%%%%%%%%%%%%%%%%%%%%%%%%%%%%%%%%%%%%%%%%%%%%%%%%%%%%%%%%%%%%%%%%%%%%%%%%%%%%%%%

\section{Result} \label{sec:result}
Here we present our SL mass reconstruction results
for $Ares$, $Hera$, and A1689. 
In \textsection\ref{subsec:result_mock_clusters}, we discuss the results for the two mock clusters and compare them with the ``truth". The A1689 result based on the multiple images in the gold sample is described in \textsection\ref{subsec:result_a1689}. We present the results for difference choices of multiple images in Appendix \ref{sec:result_a1689_etc}.
The plate scales are 5.998 kpc$/\arcsec$ at the redshift ($z=0.5$) of $Ares$, 6.043 kpc$/\arcsec$ at the redshift ($z=0.507$) of $Hera$, and 2.963 kpc$/\arcsec$ at the redshift ($z=0.18$) of A1689.

%%%%%%%%%%%%%%%%%%%%%%%%%%%%%%%%%%%%%%%%%%%%%%%%%%%%%%%%%%%%%%%%%%%%%%%%%%%%%%%%%%%%%%%%%

\subsection{Mock Cluster Result} \label{subsec:result_mock_clusters}

\begin{figure*}
\centering
\includegraphics[width=0.95\textwidth]{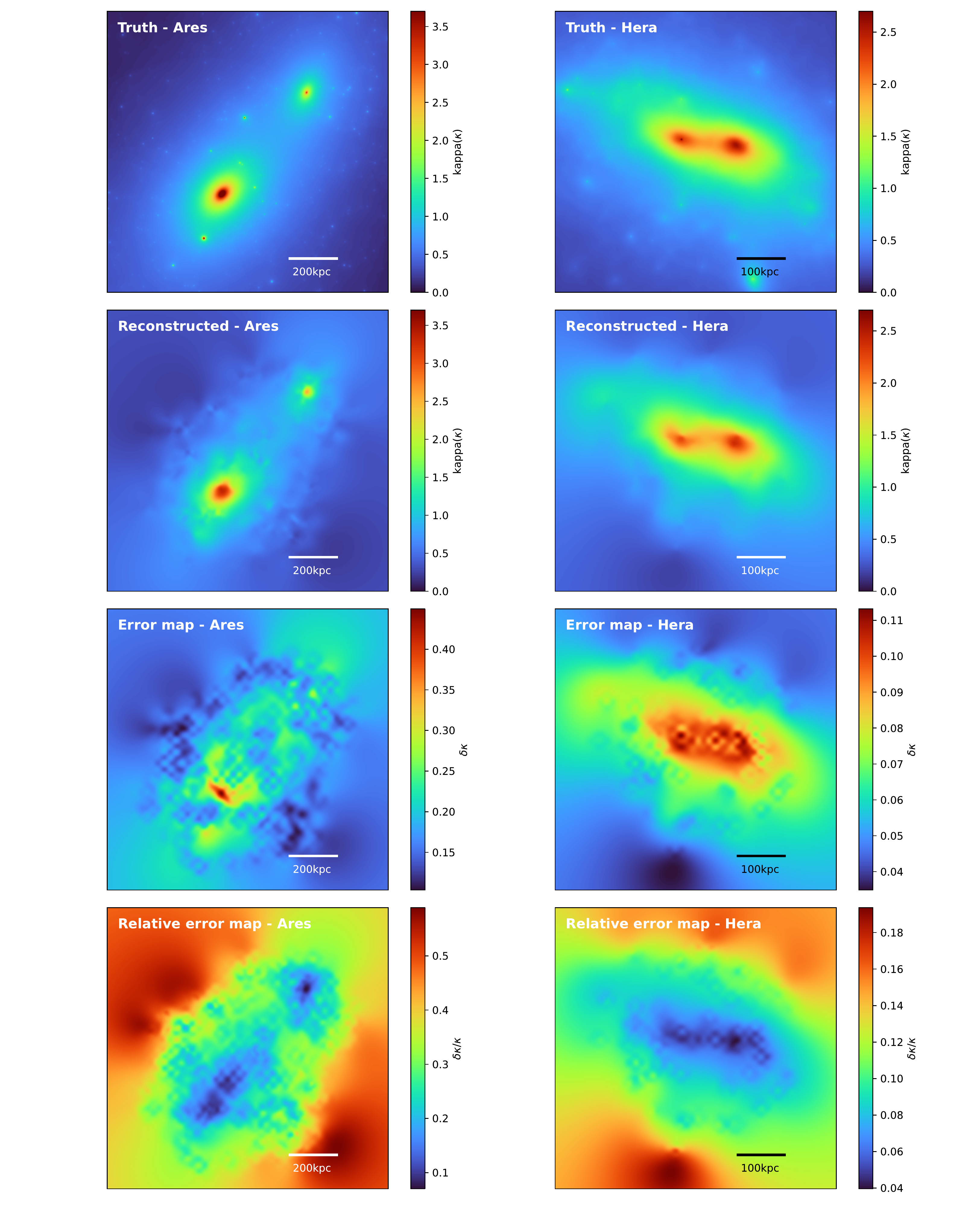} 
\caption{Mass reconstruction of $Ares$ and $Hera$ with {\tt MARS}.
Top: true $\kappa$ maps. Second row: reconstructed $\kappa$ maps from {\tt MARS}. Third row: absolute error maps of the reconstructions. Bottom: relative error maps of the reconstructions. The reference redshift is $z_{f}=9$ here and throughout the paper.}
\label{mock_truth_kappa}
\end{figure*}

\begin{figure*}
\includegraphics[width=\textwidth]{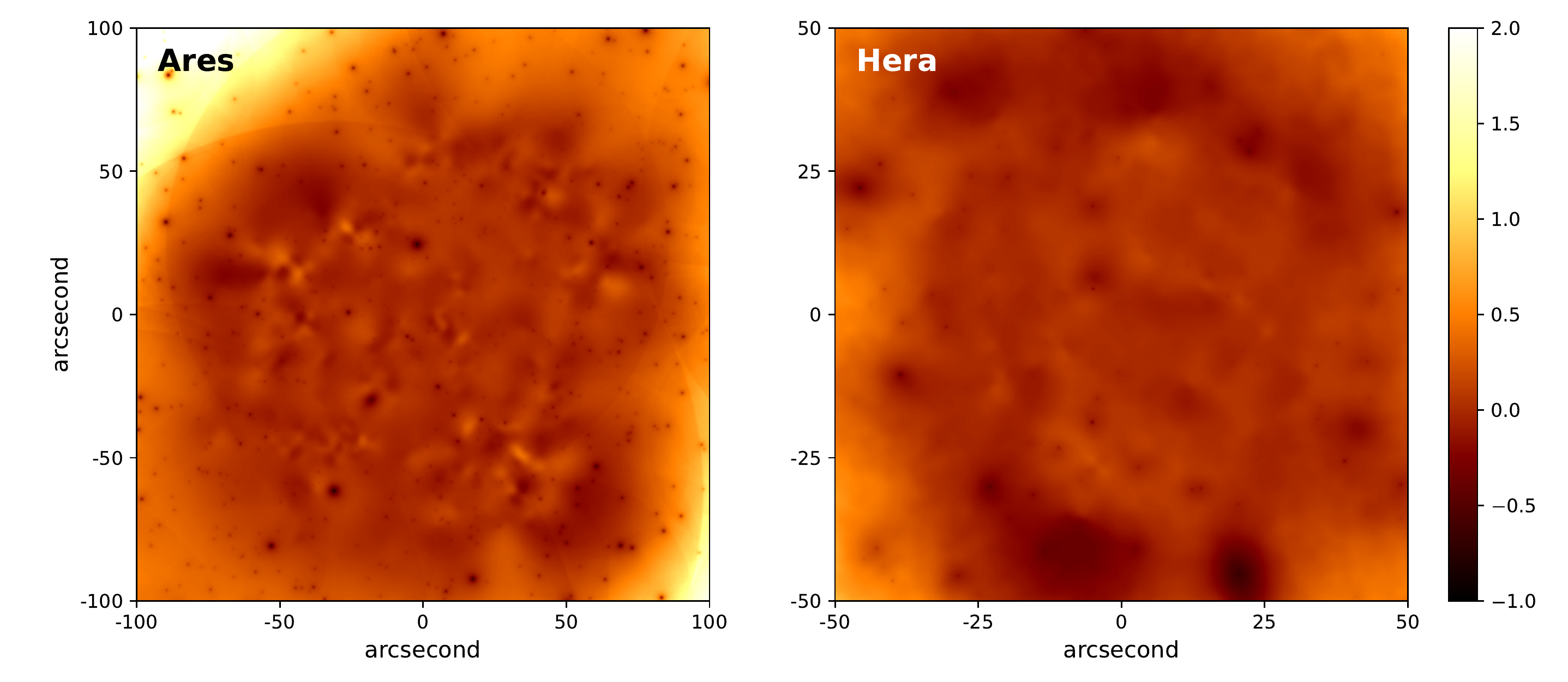}
\caption{Difference between the truth and our reconstruction.
We display the residual $(\kappa-\kappa_{truth})/\kappa_{truth}$
with the same color scheme and range used in M17 for easy comparison (see text for detail).
}
\label{fig:mock_relative_differ}
\end{figure*}

\begin{figure*}
\centering
\includegraphics[width=\textwidth]{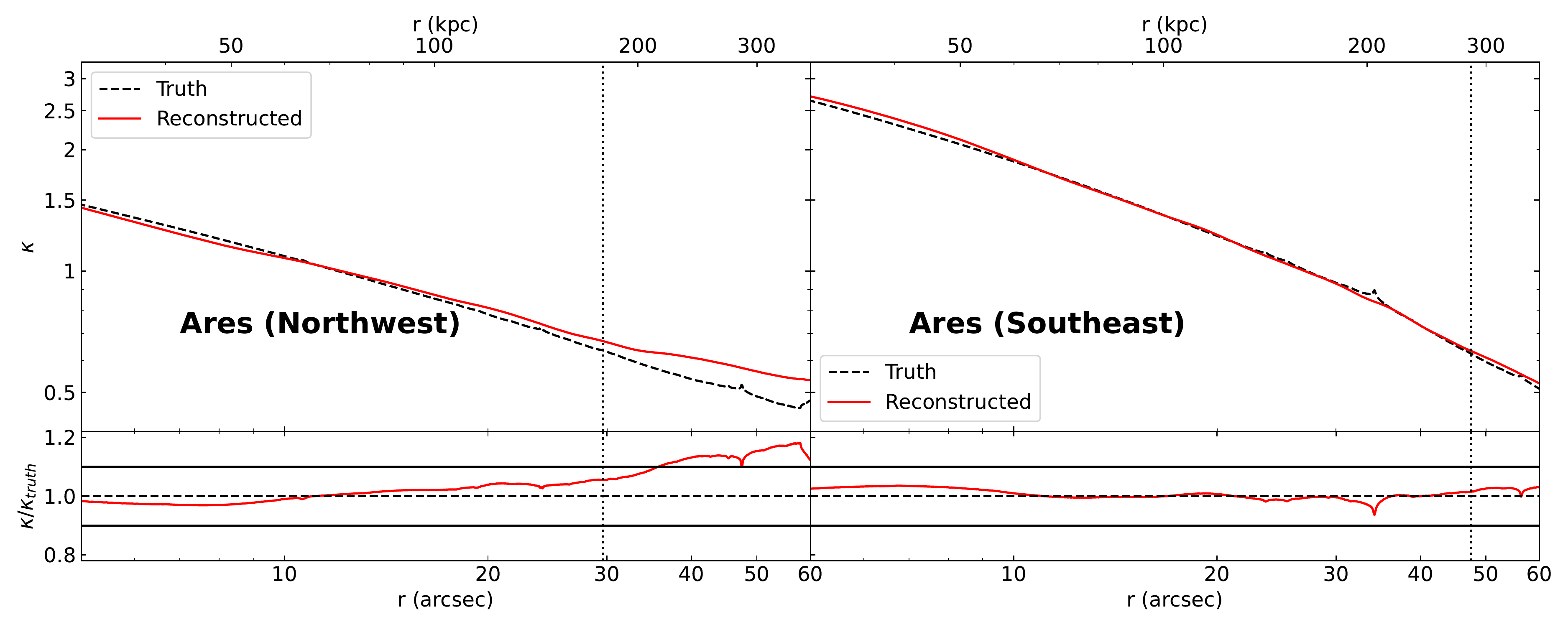} 
\caption{Radial convergence profiles of $Ares$. Bottom panels show the ratio of the reconstructed to the truth convergence profiles. The horizontal black solid lines indicate the $\pm10\%$ departures. The vertical dotted lines correspond to the approximate locations of the Einstein radii, inside which the SL data provide meaningful constraints.}
\label{radial_kappa_ares}
\end{figure*}

\begin{figure*}
\centering
\includegraphics[width=\textwidth]{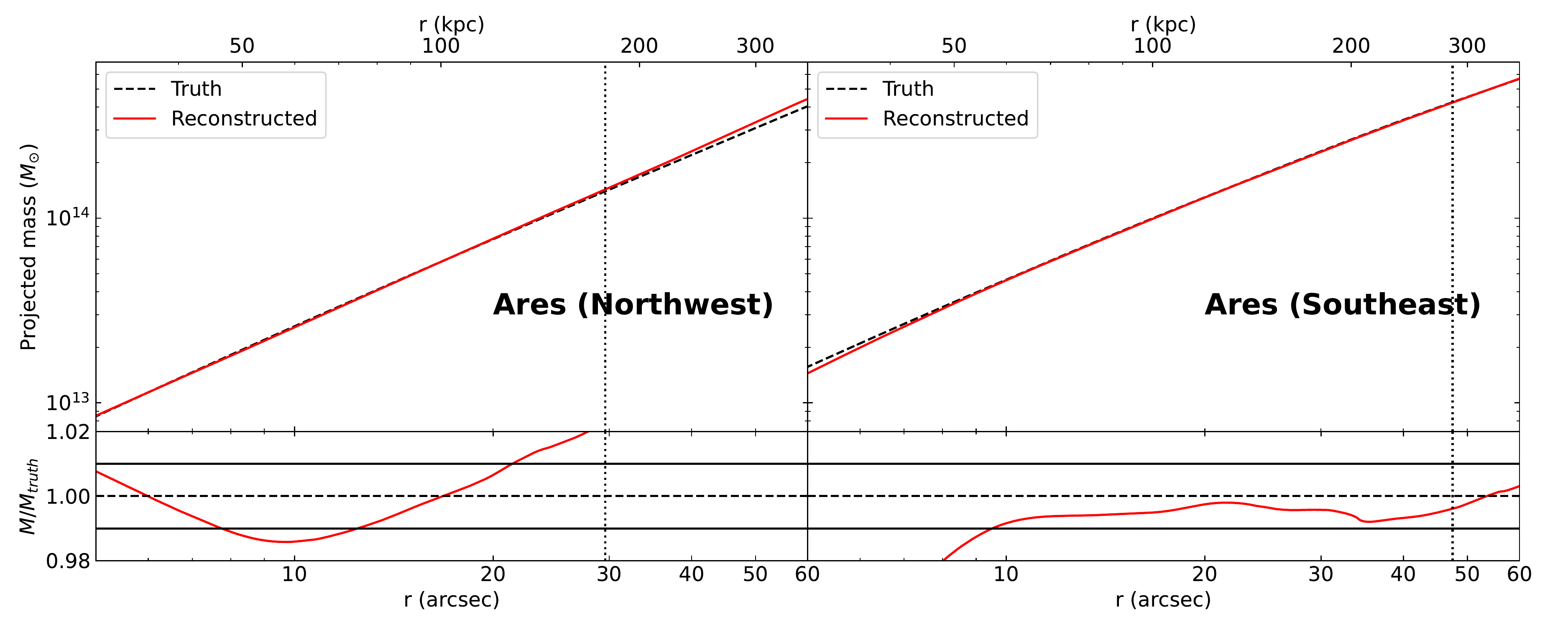} 
\caption{Cumulative projected mass profiles of $Ares$. The rest are the same as the descriptions in Figure~\ref{radial_kappa_ares} except that the horizontal black solid lines indicate $\pm1\%$ departures.}
\label{cumulative_mass_ares}
\end{figure*}

\begin{figure*}
\centering
\includegraphics[width=\textwidth]{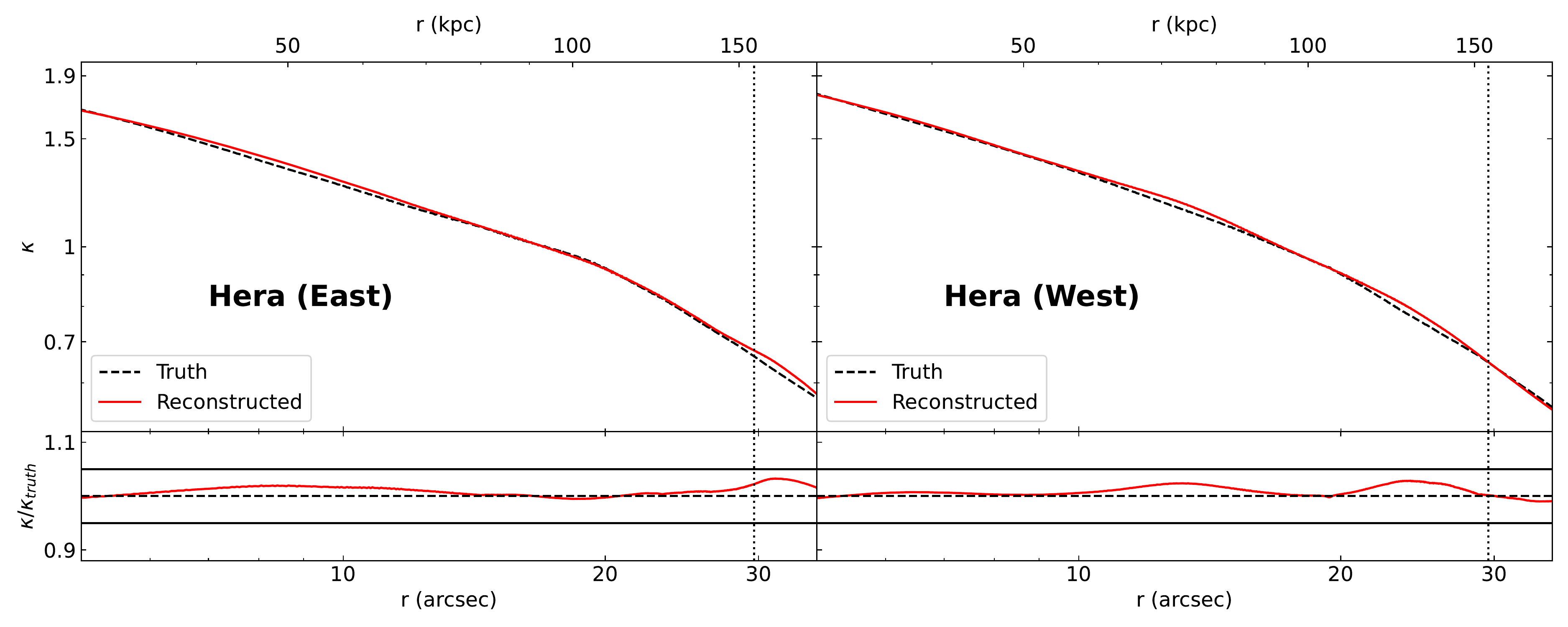} 
\caption{Radial convergence profiles of $Ares$. Same as Figure~\ref{radial_kappa_ares} except that the black solid lines in the bottom panels indicate the $\pm5\%$ differences.}
\label{radial_kappa_hera}
\end{figure*}

\begin{figure*}
\centering
\includegraphics[width=\textwidth]{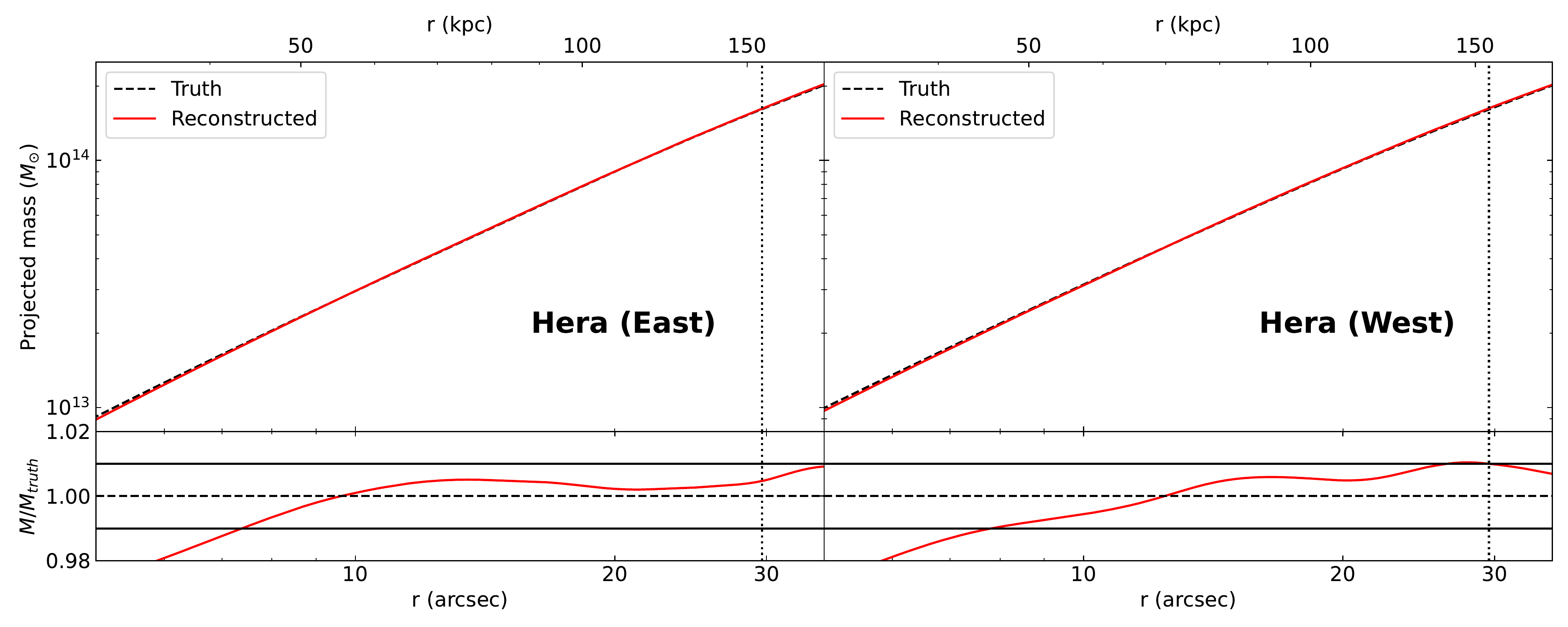} 
\caption{Cumulative projected mass profiles of $Hera$. The rest are the same as the descriptions in Figure~\ref{radial_kappa_hera} except that the horizontal black solid lines indicate $\pm1\%$ departures.}
\label{cumulative_mass_hera}
\end{figure*}

\begin{figure*}
\includegraphics[width=\textwidth]{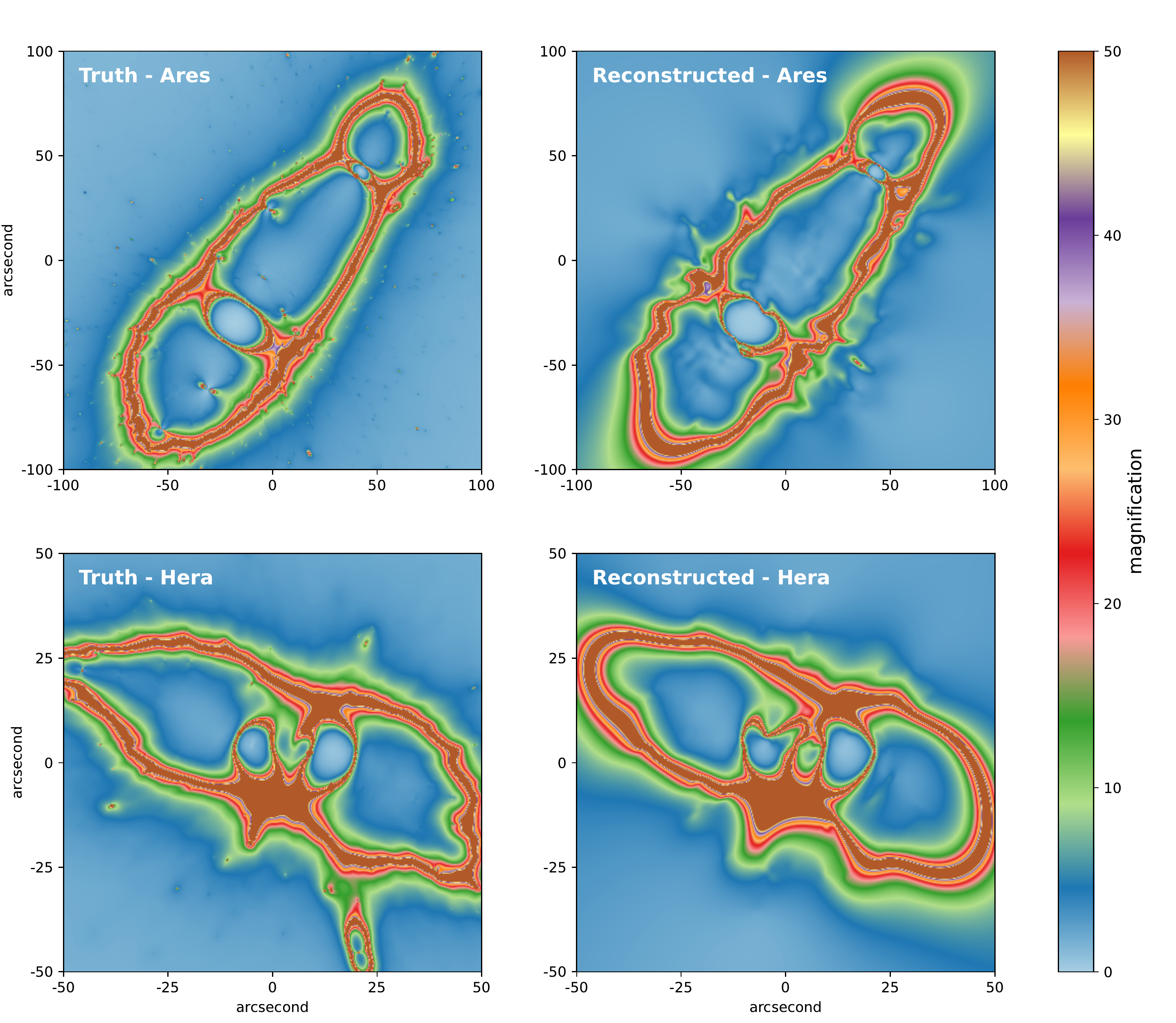}
\caption{Magnification map of the mock clusters. Left panels (right panels) show the truth (reconstructed) magnification maps. We use the same color scheme and range used in M17 for easy comparison.}
\label{fig:mock_magnification}
\end{figure*}

\begin{figure*}
\includegraphics[width=\textwidth]{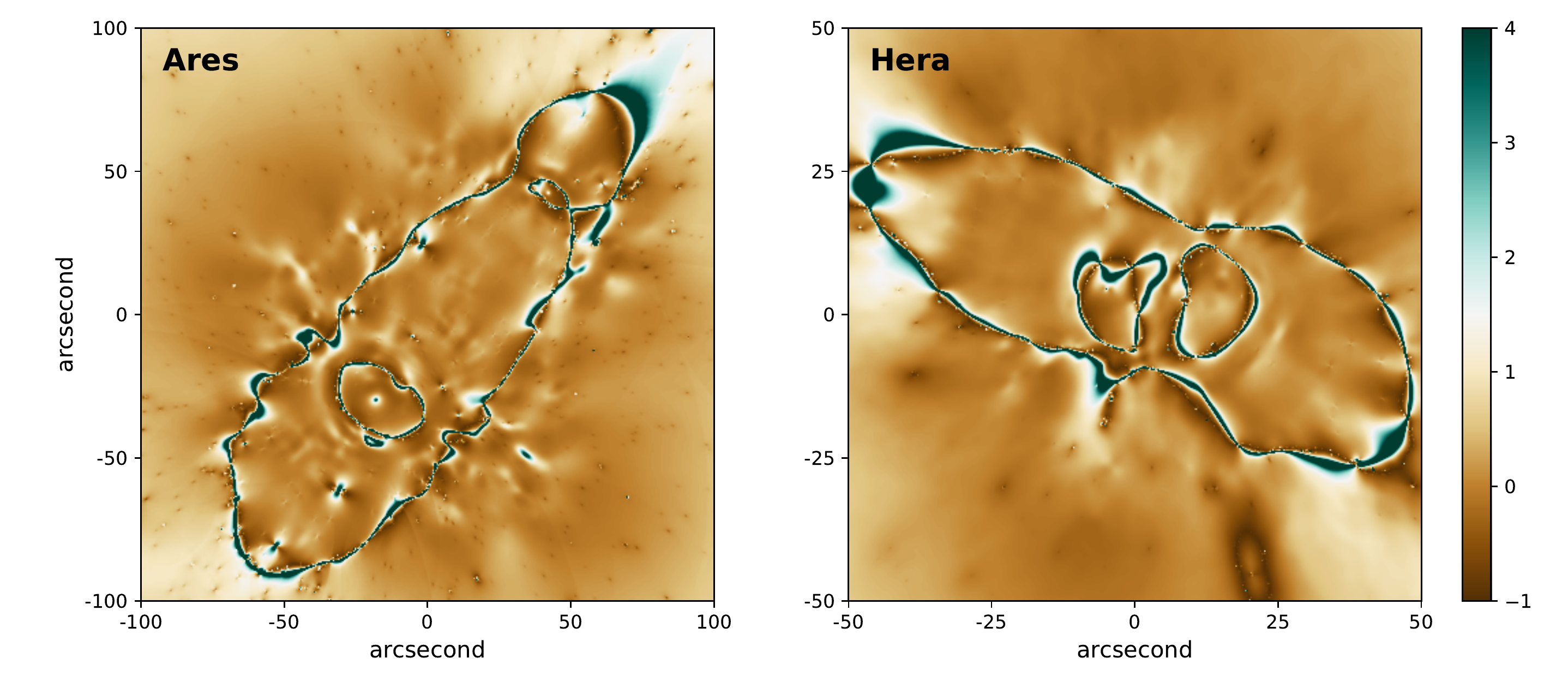}
\caption{Relative difference between the reconstructed and true magnification maps. The left (right) panel shows the result for $Ares$ ($Hera$). We use the same color scheme and range used in M17 for easy comparison.}
\label{fig:mock_magnification_residual}
\end{figure*}

\begin{figure*}
\includegraphics[width=\textwidth]{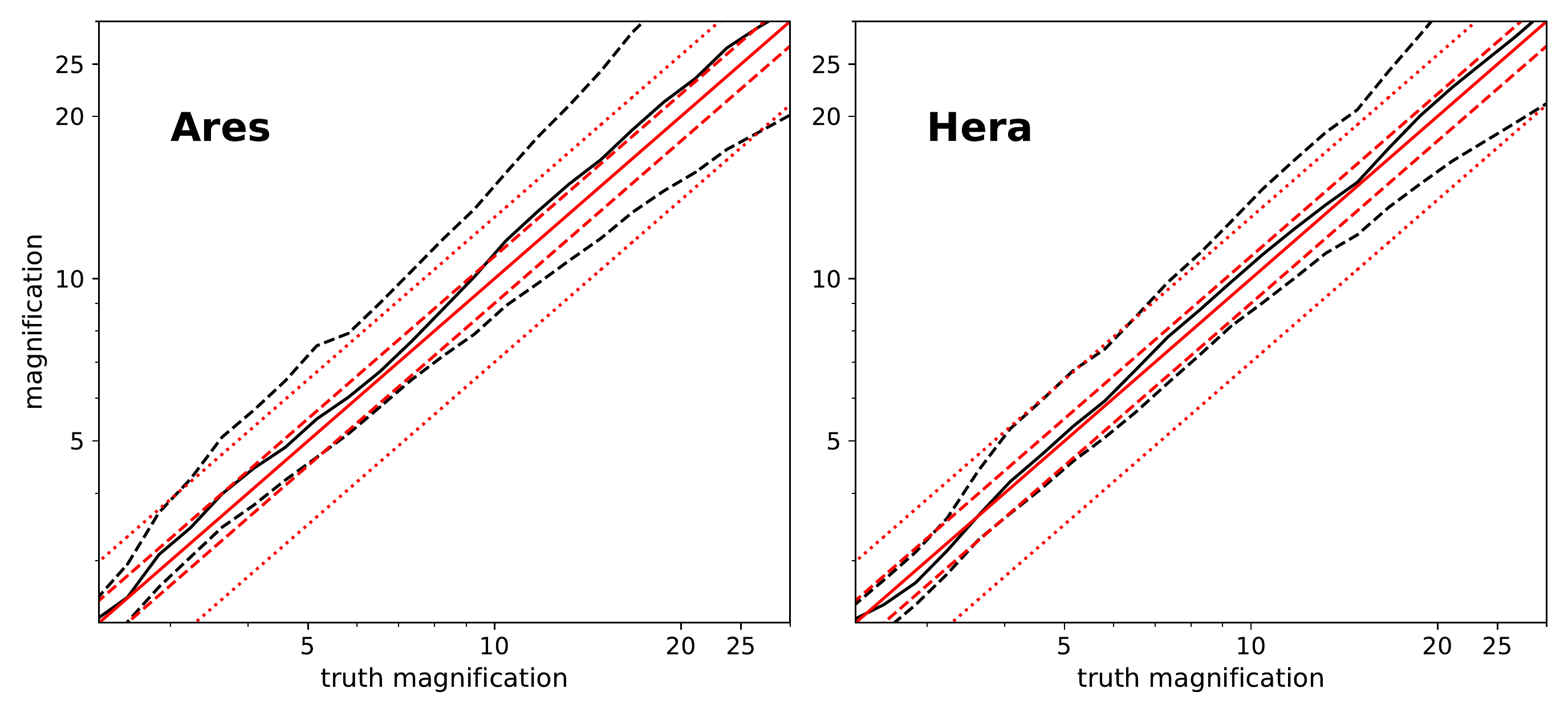}
\caption{Correlation between the reconstructed and true magnification values. The black solid lines indicate the median and the black dashed lines show the 25th and 75th percentiles. The red solid lines mark the perfect correlation between the reconstructed and true magnification. The red dashed (dotted) lines correspond to $\pm10\%$ ($\pm30\%$) differences from the truth.}
\label{fig:mock_magnification_correlation}
\end{figure*}

\subsubsection{Convergence Map Comparison}
Figure~\ref{mock_truth_kappa} shows the truth, reconstructed convergence, and uncertainty maps of $Ares$ and $Hera$. Our reconstruction closely traces the main clumps of the truth. 
In the case of $Ares$, the reconstructed convergence map reveals the two main clumps located in the NW and SE regions. The small galaxy-scale spikes are missing in our mass reconstruction because their features are smaller than our gird cell. Instead, our convergence shows small ``wrinkles", which we believe are artifacts due to the resolution limit (for example, it is difficult to make two multiple images within a grid cell converge in the source plane).

The reconstructed mass map of $Hera$ also closely follows the truth for the main clumps.
We find that the artifacts mentioned for the $Ares$
reconstruction are substantially reduced here because the true mass distribution, on which the SL data are based, is much smoother. 
In $Hera$, one noticeable discrepancy between our reconstruction and the truth is the location of the eastern peak,
which is shifted south by $\mytilde10$~kpc with respect to the truth. We attribute the offset to the combined effect of the relatively insufficient number of multiple images around this eastern peak and the regularization.

Our uncertainty map shows that the uncertainty varies from $\sigma_{\kappa}\sim0.1$ (0.035) to $\mytilde0.45$ (0.11) for $Ares$ ($Hera$). The uncertainty distribution is consistent with our expectation that in general the error is large where the convergence is high. Also, the error is somewhat reduced at the location of the multiple images. 

M17 present residual [$(\kappa-\kappa_{true})/\kappa_{true}$]
convergence maps for the nine different SL mass reconstruction algorithms, which participated in their SL modeling comparison project.
M17 show that in general the LTM models perform better for both $Ares$ and $Hera$. This is somewhat expected 1) because the two mock clusters were created under the LTM assumption and 2) because analytic profiles are reasonably good priors in the outer regions where there are no multiple images.
In M17, the performance difference between the LTM and free-form methods is greater for $Ares$, which is comprised of cuspy parametric halos.
For comparison with the results in M17, we also display our residual convergence maps with the same color table and range in Figure~\ref{fig:mock_relative_differ}. The residual map for $Ares$ shows that our mass reconstruction quality 
is similar to those of the LTM models in the region where the SL data are present. Compared with {\tt GRALE}, which leads the free-form method in performance in M17, our mass reconstruction gives smaller residual convergence values over a larger area with less large scale features.
In the case of $Hera$ (right panel of Figure~\ref{fig:mock_relative_differ}), our reconstruction is superior to any of the M17 results in terms of the amplitude and uniformity of the residuals (see Figure 10 in M17). All LTM residuals possess galaxy-scale spikes, which we attribute to mismatches between the assumed and true halo shapes.
All free-form methods suffer from large-scale residuals.

As for the mass estimation of galaxy-scale substructures, MARS cannot constrain them because the current grid-based representation cannot distinguish the projected mass belonging to the cluster halo from those bound to the substructures. In addition, these galaxy-scale substructures are smaller than our grid cell. M17 shows that some LTM-based methods outperform free-from algorithms for the substructure mass recovery, which is not surprising when we consider the fact that the LTM-methods explicitly include the galaxy-scale halos into their models.

\subsubsection{Radial Profile Comparison }
We compare the radial convergence and cumulative mass profiles in Figures~\ref{radial_kappa_ares},~\ref{cumulative_mass_ares},~\ref{radial_kappa_hera}, and \ref{cumulative_mass_hera}. 
M17 state that the SL data are expected to provide tight constraints at $20\arcsec < r< 60\arcsec$ ($10\arcsec < r< 30\arcsec$) for $Ares$ ($Hera$). Thus, we choose the radial ranges for comparison accordingly.
Since both $Ares$ and $Hera$ are comprised of two main halos, we measure the radial profiles twice for each cluster using the two centers of the main clumps. Note that M17 use only the stronger peak (the southeastern peak for $Ares$ and the western peak for $Hera$) for each cluster to measure the profiles.
Overall, both radial convergence and cumulative mass profiles are in excellent agreement with the truth. In the study of M17, the three best-performing LTM methods produce the mass profiles for $Ares$, which differ from the truth by $<2$\%, while most free-form methods depart from the truth by $5-15$\%.
Our result (see the right panel of Figure~\ref{cumulative_mass_ares}) shows that the agreement of the cumulative mass profile in the $20\arcsec < r< 60\arcsec$ radial range is at the $\lesssim1$\% level.
When we choose the northwestern peak as the reference (note that this measurement is not presented in M17), the difference starts to rapidly increase at $r>30$ and reaches $\mytilde10\%$ at $r=60\arcsec$ (see the left panel of Figure~\ref{cumulative_mass_ares}). Since the northeastern peak has a smaller ($<\mytilde30\arcsec$) Einstein radius, this behavior is not surprising.
In the case of $Hera$, the mass profiles agree with the truth within $\lesssim1$\% for the radial range $10\arcsec < r< 30\arcsec$ regardless of the center choice. The best LTM-methods in M17 show agreements at the $\mytilde3$\% levels. The free-form results in M17 differ from the truth by 5-10\%.

\subsubsection{Magnification Comparison }
In Figure~\ref{fig:mock_magnification}, we display the magnification maps for $Ares$ and $Hera$. Also shown in Figure~\ref{fig:mock_magnification} are the true magnification maps that we computed from the publicly available convergence and shear data.
The overall shape and size of the reconstructed critical curves are well-recovered, although the largest discrepancies in magnification are found near the
critical curves as shown in the residual maps (Figure~\ref{fig:mock_magnification_residual}). As noted by M17, the critical curves define the regions where the magnification diverges and thus small differences in the mass reconstruction can cause large magnification errors near them.
M17 report that some free-form approach such as the Bradac-Hoag, Coe, and Diego-multire methods produce highly fluctuating
irregular critical lines, which are not observed in the truth. As shown, our entropy-based regularization enables us to obtain relatively smooth critical curves. Following M17, we also compute the correlation between our and truth magnification maps and display the result in Figure~\ref{fig:mock_magnification_correlation}, which indicate that the accuracy (i.e., departure of the median from the perfect correlation) is within $\mytilde10$\% in the range $1<\mu<30$ whereas the precision (i.e., distance between the 25th and 75th percentiles) is $\mytilde20$\% at $\mu=10$. Note that some details in our magnification comparison methods 
may be different from M17. For example, we do not use exactly the same region for comparison. 
With this caveat, we report that our prediction on the median magnification is accurate within ~10\% for both Ares and Hera. In the case of Ares, we report that our magnification reconstruction results are significantly better than some (Bradac-Hoag, Diego-multires, Lam, and Coe) of the non-LTM methods while the Diego-reggrid and GRALE methods show a similar performance to ours. Clearly, most LTM-methods outperform MARS for the Ares magnification reconstruction. In the case of Hera, the performance of MARS is similar to those of GLAFIC and Diego-overfit, which are the best-performing methods in the LTM and free-form algorithms, respectively.
Since the two mock clusters are generated under the LTM assumption, there exists a strong correlation between dark matter and luminous galaxies in the mock data. As noted by M17, it is possible that LTM approaches can benefit from the use of this extra information. M17 show that most LTM methods perform better in the case of Ares than Hera, which may hint at this possibility because Ares has tighter dark matter-galaxy correlations.
In addition, the critical curves from the LTM methods are much smoother by construction, which is a significant advantage because even small irregular features around the critical curves in non-LTM methods critically contribute to the degradation of the magnification correlations.

%%%%%%%%%%%%%%%%%%%%%%%%%%%%%%%%%%%%%%%%%%%%%%%%%%%%%%%%%%%%%%%%%%%%%%%%%%%%%%%%%%%%%%%%%

\subsection{A1689 Result} \label{subsec:result_a1689}

Here we focus mainly on the convergence, uncertainty, magnification, and radial profile. In Appendix, we present our reconstruction of the multiple images in the source plane and the giant arcs in the image plane.

\begin{figure}
    \plotone{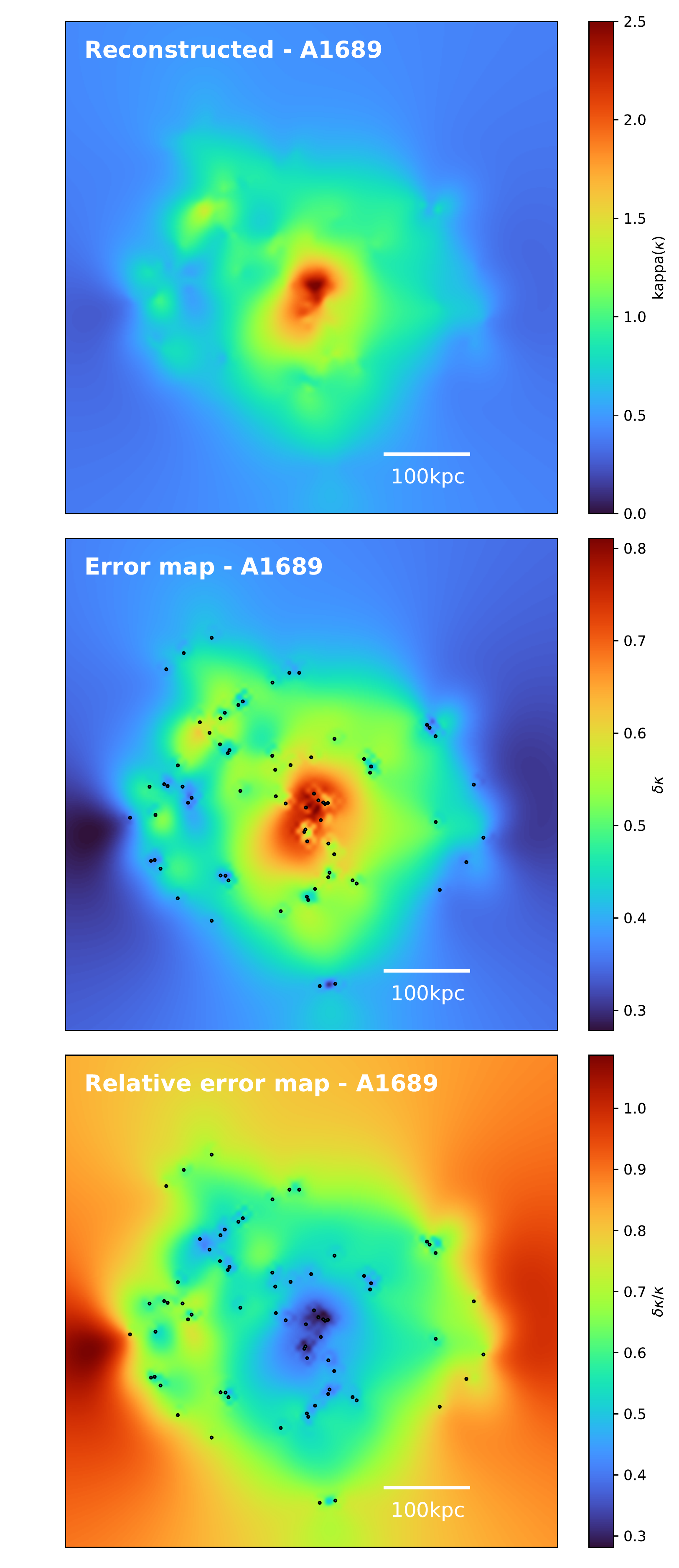}
    \caption{A1689 mass reconstruction with {\tt MARS}. Top panel shows the $\kappa$ distribution within the $200\arcsec\times200\arcsec$ target reconstruction field. In the middle panel, we display the uncertainty ($\delta \kappa$) map of the mass reconstruction utilizing the Hessian matrix. In the bottom panel, we plot the relative error ($\delta \kappa/\kappa$) map. The black dots indicate the positions of the multiple images.
    } 
    \label{fig:result_A1689}
\end{figure}

\begin{figure}
    \plotone{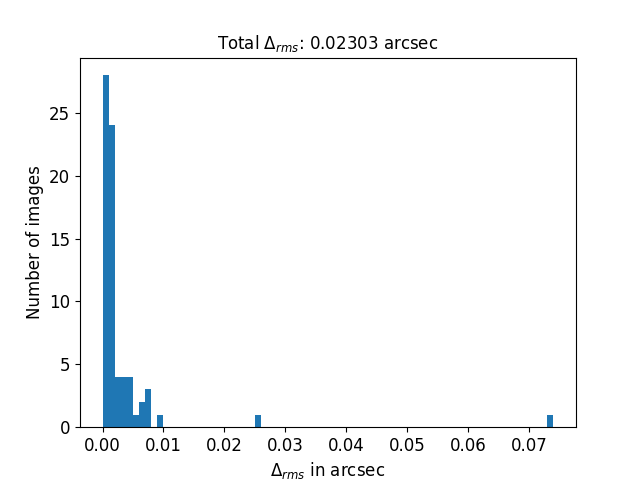}
    \caption{Multiple image scatters in the lens plane of the A1689 field. Except for the two relatively large scatters at $\mytilde0.025\arcsec$ and $\mytilde0.073\arcsec$, most multiple images have negligible ($<0.01\arcsec$ lens plane scatter.
    Note that these results are derived after excluding false multiple images (see text).} 
    \label{fig:lens_rms_a1689}
\end{figure}

\subsubsection{Convergence and Uncertainty Maps} \label{sec:convergence}
\begin{figure*}
\centering
\includegraphics[width=0.9\textwidth]{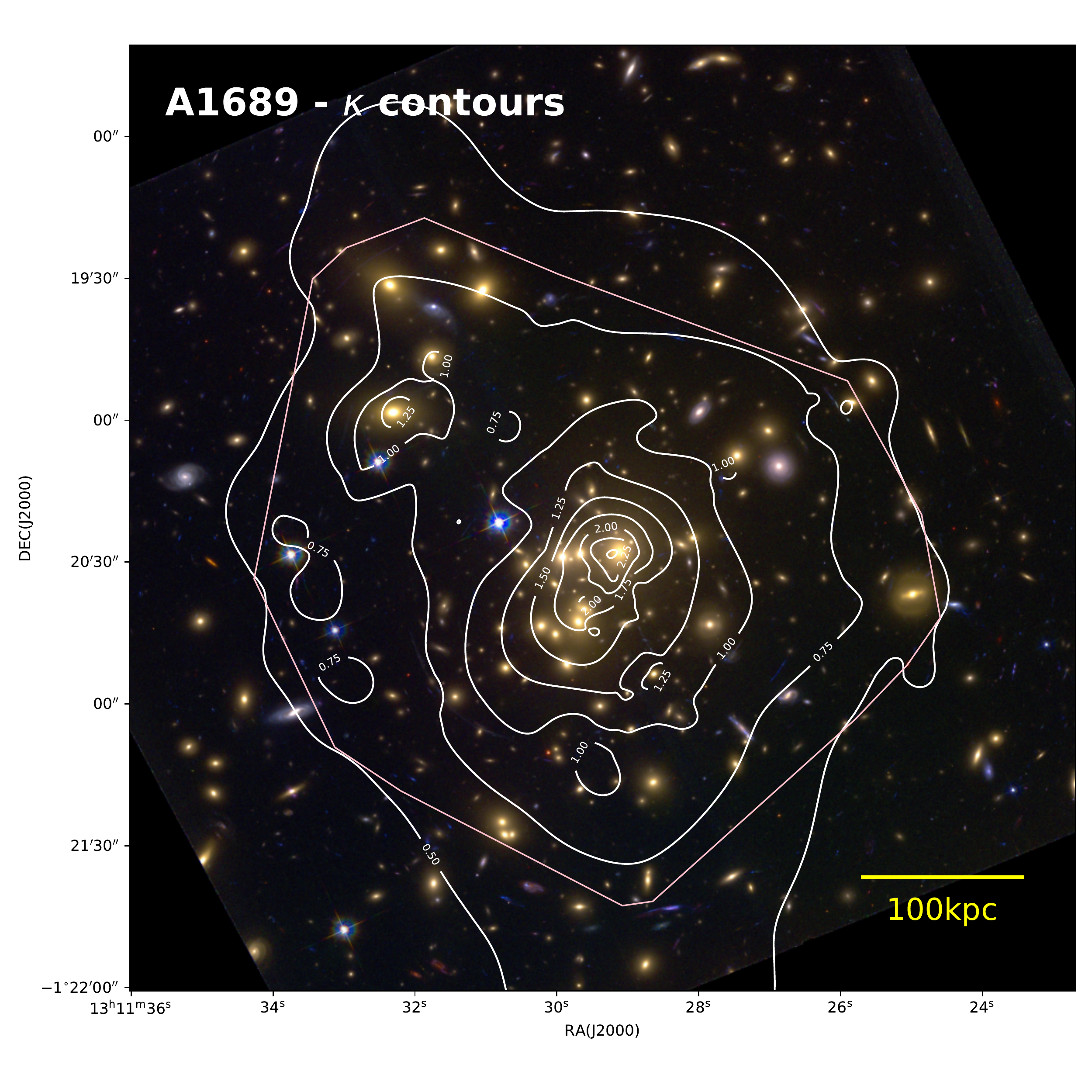}
\caption{Mass contours overlaid on the ACS color-composite image. The white contours and labels show the convergences. The convex hull (solid pink) encloses the area where multiple images are located. The color-composite image is
created with the F475W (blue), F625W (green), and F775W (red) filters.
The projected mass distribution of A1689 is characterized by the primary mass peak centered on the BCG at $(\alpha,\delta)\sim(13^h11^m29^s,-1\degr 20\arcmin 30\arcsec)$ and the much weaker secondary mass peak centered on the bright galaxy at $(\alpha,\delta)\sim(13^h11^m32^s,-1\degr 20\arcmin 00\arcsec)$. 
}
\label{fig:result_A1689_gold_real}
\end{figure*}

We display the reconstructed $\kappa$ distribution and uncertainty map of A1689 in Figure~\ref{fig:result_A1689}. Also, Figure~\ref{fig:result_A1689_gold_real} shows the $\kappa$ contour overlaid on the color-composite image of the cluster.
Readers are reminded that here we only present the result obtained with the gold sample. We discuss how the result changes when we include the gold candidate or silver samples in Appendix~\ref{sec:result_a1689_etc}.

The mass map shows that the projected mass distribution of A1689 is characterized by the dominant mass peak centered on the BCG at $(\alpha,\delta)\sim(13^h11^m29^s,-1\degr 20\arcmin 30\arcsec)$ and the secondary, but much weaker, mass peak centered on the bright galaxy at $(\alpha,\delta)\sim(13^h11^m32^s,-1\degr 20\arcmin 00\arcsec)$. This feature has also been observed in some previous LTM-based studies \citep[e.g.,][]{2005ApJ...621...53B,2006ApJ...640..639Z,2006MNRAS.372.1425H,2007ApJ...668..643L}.
Some free-form methods do not constrain the secondary peak mentioned above as the characteristic feature of A1689. For example, \citet[]{2010ApJ...723.1678C} obtained a mass map with several substructures whose projected densities are comparable and do not identify our secondary clump as a distinguishing feature of A1689. \citet[]{2005MNRAS.362.1247D} presented a result from averaging over 1000 solutions. Although their result shows a weak secondary substructure near our mass peak, it is offset by $\mytilde20\arcsec$, which is surprising if true within the LTM paradigm. We note that in their revised analysis \citep[]{2015MNRAS.446..683D}, this secondary peak coincides with the galaxy.

Given that our mass map is the solution that successfully delens all multiple images to converge in the source plane with a negligible scatter of $\mytilde0.001\arcsec$, its smoothness is remarkable for a free-from result. As already mentioned, the mass reconstructions from \citet[]{2010ApJ...723.1678C} and \citet[]{2005MNRAS.362.1247D} suffer from high-amplitude small-scale fluctuations.
\citet[]{2014MNRAS.439.2651M} produce a smooth mass reconstruction of A1689 using the GRALE method, which does not assume LTM.
However, since the authors still assume that the cluster consists of Plummer spheres, this smoothness is achieved by design.

We also present the multiple image scatters in the image plane in Figure~\ref{fig:lens_rms_a1689}. Except for the two largest scatters at $\mytilde0.025\arcsec$ and $\mytilde0.073\arcsec$, we find that most multiple images in the A1689 field have a small ($<0.01\arcsec$) scatter.
We note that these scatters are evaluated without including false multiple images, which would produce catastrophically large scatters. Since MARS is designed to minimize scatters only in the source plane, the model cannot prevent false multiple images. This weakness is shared by the SL reconstruction algorithms that employ only source plane minimization.

Our scrutiny of the mass map indicates that the mass map contains several ``pinches" or ``quadrupoles" characterized by alternating over- and under-dense regions. We believe that they are artifacts caused by the resolution limitation and happen where compact halos generate multiple images densely distributed in small areas. As discussed in \textsection\ref{subsec:result_mock_clusters}, the mass reconstruction for $Ares$ produces many such features while no such artifact is seen in the result for $Hera$. 
The uncertainty map (middle panel of Figure~\ref{fig:result_A1689}) somewhat resembles the mass reconstruction (top panel of Figure~\ref{fig:result_A1689}), which is similar to our mock cluster cases (\textsection\ref{subsec:result_mock_clusters}). 
Near the central mass peak, this convergence uncertainty ($\delta \kappa$) is high ($\delta \kappa\sim0.8$) whereas near the field boundary it is low ($0.3\lesssim \delta \kappa \lesssim 0.4$).
When rescaled by the local convergence, the relative ($\delta \kappa/\kappa$) error map (bottom panel) shows the opposite trend with the estimate being high ($\delta \kappa/\kappa\sim1$) near the field edges, which is consistent with the relative difference map shown for the mock cluster (Figure~\ref{fig:mock_relative_differ}). Note that the uncertainties are relatively low at the location of the multiple images.

\begin{figure}
\centering
\includegraphics[width=0.48\textwidth]{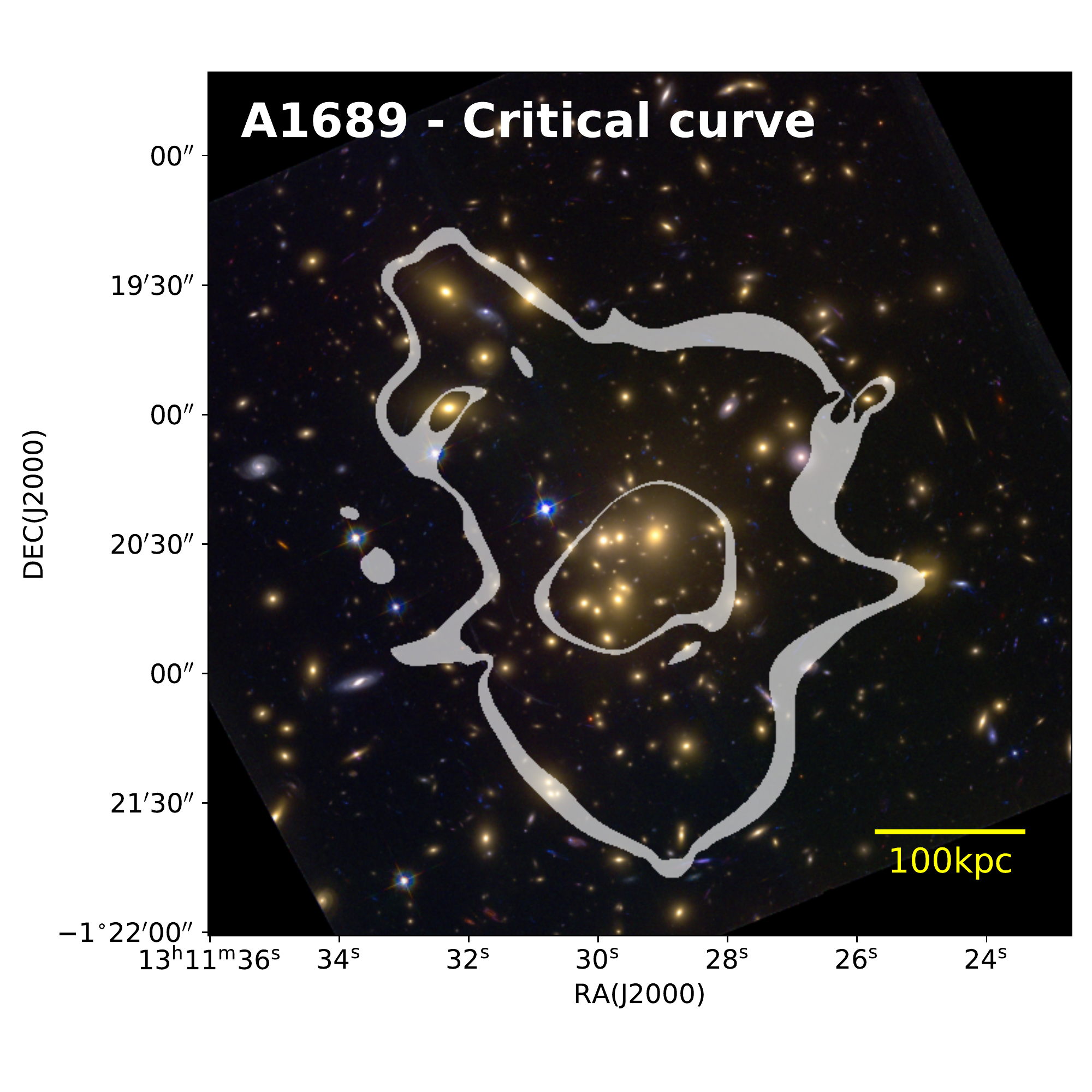}
\caption{Critical curve of A1689. The white curves indicate the regions where the magnification is greater than 50.  The overall morphology
comprised of the inner (radial) $r\sim15\arcsec$ and outer (tangential) $r\sim40\arcsec$ critical curves
is similar to the findings in previous LTM studies.
}
\label{fig:result_A1689_critical_curve}
\end{figure}

\subsubsection{A1689 Magnification}\label{a1689_magnification}
Figure~\ref{fig:result_A1689_critical_curve} presents our critical curves of A1689. The global morphology
comprised of the inner (radial) $r\sim15\arcsec$ and outer (tangential) $r\sim40\arcsec$ critical curves
is similar to the findings in previous studies. However, details differ greatly among the studies including ours. Therefore, despite the fact that A1689 is one of the most studied SL clusters, the community still needs to converge on the magnification details and thus the interpretation of high-$z$ galaxies found behind the cluster.
We repeat that since the relation between mass and magnification distributions are highly nonlinear, even small differences in the convergence lead to large changes in the magnification.

As seen in the mock cluster results, the smoothness of our critical curve is noteworthy when compared with other free-form results. For example, since no regularization is imposed in \citet[]{2010ApJ...723.1678C}, their critical curves show highly oscillatory features and vary widely among different solutions, which however are all capable of converging all used multiple image positions in the source plane. In \citet[]{2005MNRAS.362.1247D}, only the critical curve estimated from the mean (out of 1000) convergence map is shown.

\begin{figure}
    \plotone{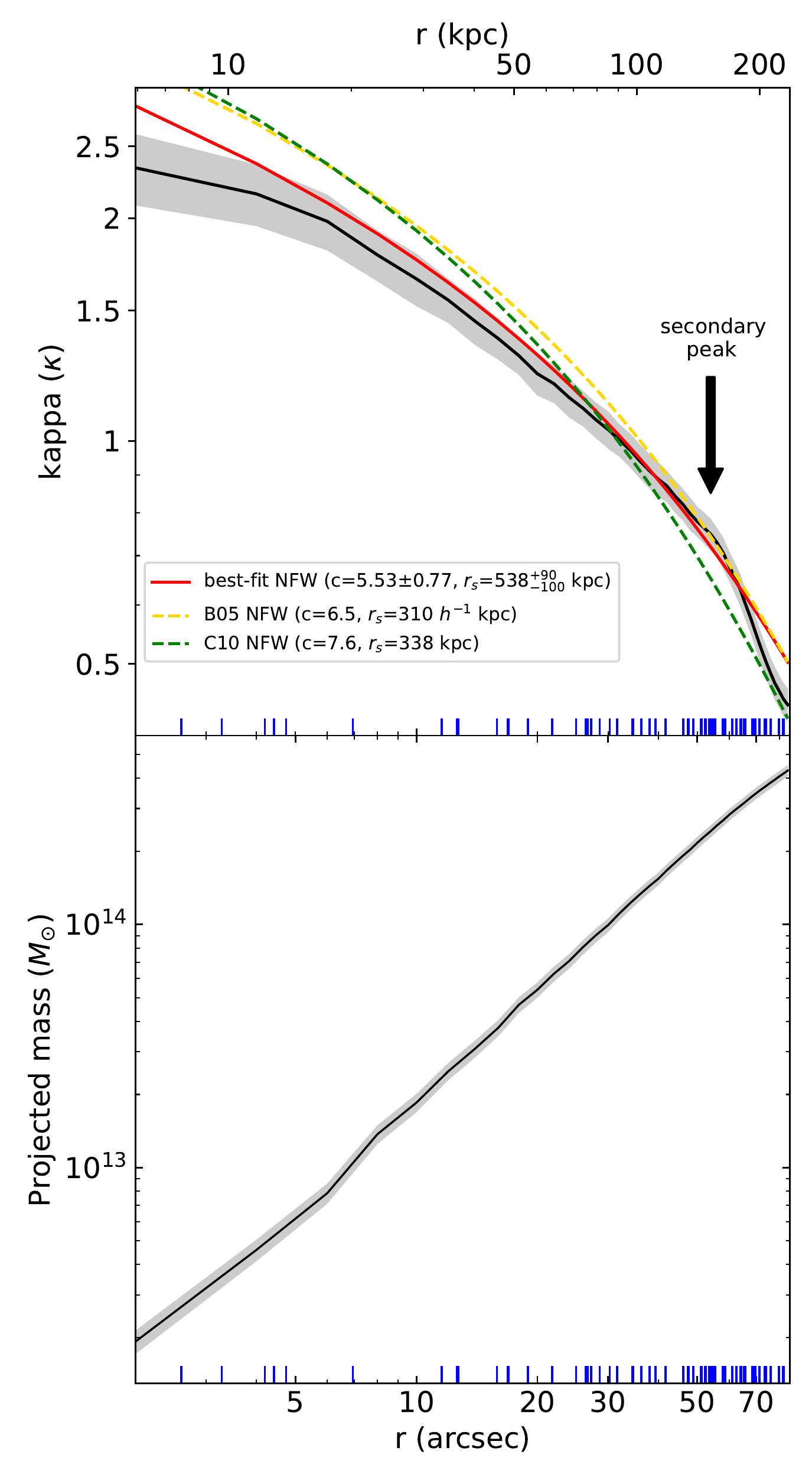}
    \caption{Radial mass profile of A1689. Blue markers indicate the locations of the multiple images. Top: Black (red) solid line shows the (best-fit NFW) radial mass density profile. The gray shade represents the uncertainty. We use the convergence values from $11.85$~kpc to 237 kpc for the NFW profile fitting and obtain $c_{200}=5.53\pm0.77$ and $r_s=538^{+90}_{-100}$ kpc. The yellow dashed line (green dashed line) indicates the NFW models fitted from B05 (C10). Bottom: We plot the projected mass profile.} 
    \label{fig:NFW_fitting}
\end{figure}

\subsubsection{A1689 Radial Profile} \label{A1689_radial_profile}
The experiments with the mock cluster SL data (\textsection\ref{subsec:mock_clusters}) show that our mass reconstruction recovers the cluster radial profiles with superb ($\lesssim1$\%) accuracy in the regions where multiple images are present.
Thus, if all multiple images are real in the gold sample, we expect that in the current study a similarly high-fidelity mass profile can be obtained for A1689 in the SL regime.

Our test with the mock cluster data shows that our mass reconstruction well-recovers the central profile down to $r\sim4\arcsec$ ($\mytilde25$~kpc).
At smaller radii, the density becomes non-negligibly underestimated because of the limitations in resolution and regularization. Considering the plate scale and resolution of our mass reconstruction for A1689, we believe that the smallest radius at which we can restore the density without severe bias is $\mytilde12$~kpc. Given the extent of the multiple images, the mass density is expected to be constrained by the data out to $\mytilde240$~kpc. 

We display the projected radial density profile of A1689 in the top panel of Figure~\ref{fig:NFW_fitting}.
Also plotted is the best-fit ($c_{200}=5.53\pm0.77$, and $r_s=538^{+90}_{-100}$~kpc) NFW profile. 
This NFW profile is a reasonable approximation of the overall shape of our radial profile in the SL regime, although we observe some indications of the densities predicted by the best-fit NFW model being $1~\sigma$ higher at $r\lesssim15$~kpc and lower at $r\gtrsim200$~kpc. Since there exist several multiple images at $r\lesssim15$~kpc, it is unlikely that the shallower (than the best-fit NFW prediction) profile is entirely due to the aforementioned artifact, but is somewhat constrained by the data. The ``shoulder" (see the arrow in the top panel) at $r\sim160$~kpc coincides with the location of the secondary mass peak at $(\alpha,\delta)=(13^h11^m32^s,-1\degr 20\arcmin 00\arcsec)$, which indicates that our best-fit NFW model might be non-negligibly affected by the presence of the substructure.
The cumulative projected mass profile (the bottom panel of 
Figure~\ref{fig:NFW_fitting}) shows that the aperture mass of A1689 is $\mytilde4\times10^{14}M_{\sun}$ at $r=237$~kpc, which is in good agreement with the previous studies \citep[e.g.,][]{2005ApJ...621...53B,2006MNRAS.372.1425H,2007ApJ...668..643L}. At face value, the best-fit concentration value $c_{200}=5.53\pm0.77$ is slightly lower than the results in some previous studies. However, when we limit our comparison to those who used only the SL data \citep[e.g.,][]{2005ApJ...621...53B, 2007ApJ...668..643L, 2010ApJ...723.1678C}, our measurement is consistent at the $\mytilde1~\sigma$ level.

%%%%%%%%%%%%%%%%%%%%%%%%%%%%%%%%%%%%%%%%%%%%%%%%%%%%%%%%%%%%%%%%%%%%%%%%%%%%%%%%%%%%%%%%%
%%%%%%%%%%%%%%%%%%%%%%%%%%%%%%%%%%%%%%%%%%%%%%%%%%%%%%%%%%%%%%%%%%%%%%%%%%%%%%%%%%%%%%%%%

\section{Discussion} \label{sec:discussion}
\subsection{Effect of Grid Resolution} \label{subsec:discussion_resolution}
\begin{figure*}
\centering
\begin{center}
    \includegraphics[width=\textwidth]{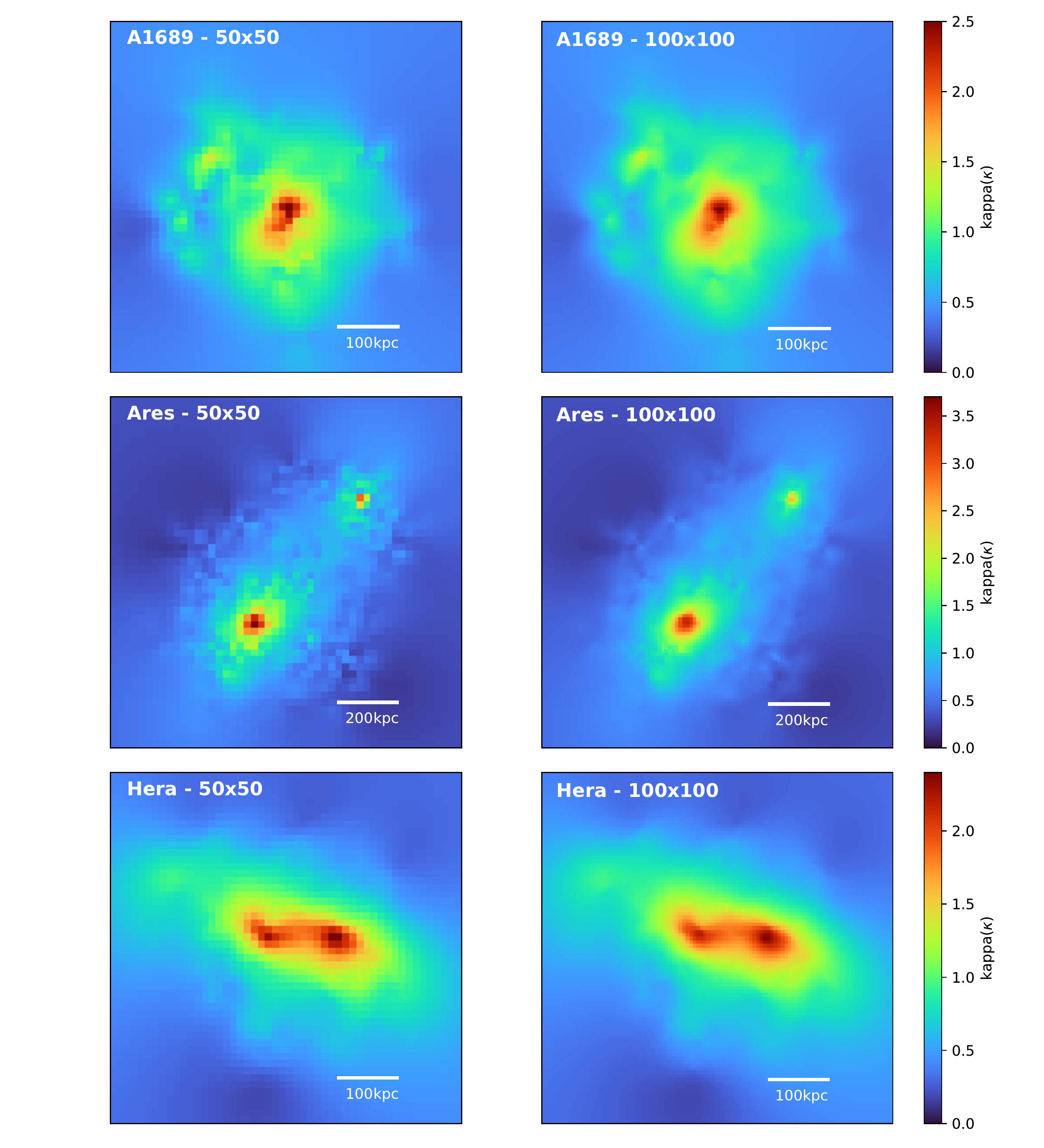}
\end{center}
\caption{Resolution test. Left panels (right panels) show the mass reconstructions with the $50\times50$ ($100\times100$) grid. 
The $100\times100$ versions
reduce quadrupole-like artifacts mentioned in \textsection\ref{sec:convergence}. In particular, the artifact reduction is most significant in $Ares$.
Apart from the artifact reduction, no significant systematic difference is found between the two resolution cases.
}
\label{resolution_comparison}
\end{figure*}

As mentioned in \textsection\ref{subsec:reconstruction_method}, our final reconstruction is achieved  in two steps (i.e., initial reconstruction in the $50\times50$ grid followed by a second reconstruction in the $100\times100$ grid\footnote{These refer to the resolutions when the margins are excluded.}). If significant systematic differences are found between the two results, this might indicate that the resolution in the second step is insufficient and the result should be further improved with even a higher resolution grid. Here we demonstrate that although the larger grid provides a higher resolution in some cases, no significant systematic difference is found.

Figure~\ref{resolution_comparison} compares the $50\times50$ and $100\times100$ versions side by side for the three clusters studies in this paper. It is clear that the larger grid improves the resolution
of the resulting mass map. In addition, the $100\times100$ versions
reduce quadrupole-like artifacts mentioned in \textsection\ref{sec:convergence}. In particular, the artifact reduction is most significant in $Ares$, whose truth map possesses many sharp spikes, which cannot be sufficiently represented even with the $100\times100$ resolution.

We also notice that the convergence values at the mass peaks are slightly reduced in the higher resolution version. We speculate that this might arise from our two-step mass reconstruction method, which obtains the lower resolution version first and uses the smoothed version as a prior to the second reconstruction. However, the differences make only a negligible impact on the radial mass profiles.

Therefore, we conclude that although the higher resolution grid improves the resolution and reduces some artifacts, the change in grid resolution does not lead to any significant systematic differences that lead to changes in scientific interpretation.

%%%%%%%%%%%%%%%%%%%%%%%%%%%%%%%%%%%%%%%%%%%%%%%%%%%%%%%%%%%%%%%%%%%%%%%%%%%%%%%%%%%%%%%%%

\subsection{Effect of Initial Conditions} \label{subsec:discussion_initial}
\begin{figure*}
\centering
\begin{center}
    \includegraphics[width=0.75\textwidth]{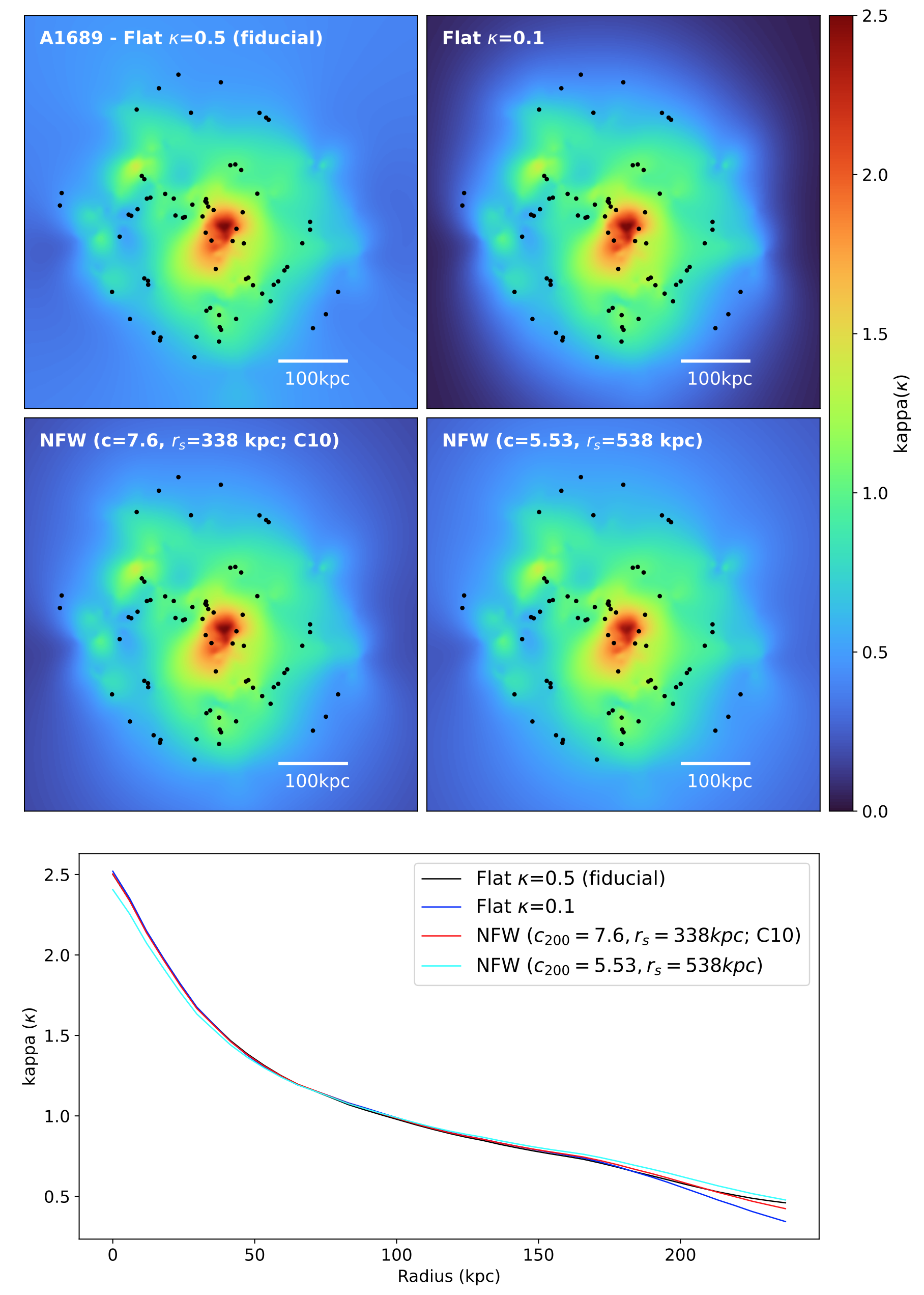}
\end{center}
\caption{Dependence of mass reconstruction on initial conditions. We experiment with four different initial $\kappa$ distributions. Upper left: a flat mass sheet of $\kappa$=0.5. Upper right: a flat mass sheet of $\kappa$=0.1. Middle left: an NFW profile from \citet[]{2010ApJ...723.1678C} ($c_{200}=7.6, r_s=338$ kpc). Middle right: an NFW profile from this work ($c_{200}=5.53, r_s=538$ kpc). 
To guide eyes, we use black dots to mark the locations of the multiple images. The mass map comparison shows that the agreement among the four results is excellent in the region constrained by the multiple images.
Bottom: comparison of the four radial profiles.
The mass map comparison shows that the agreement among the four results is excellent in the region constrained by multiple images. 
No significant difference is found out to $r\sim200$~kpc. Beyond this radius, some small systematic differences are noticeable, which is not surprising because the SL data does not provide any meaningful constraint there.
The reference redshift is $z_{f}=9$.}
\label{dif_ini_condition}
\end{figure*}

In order to show that our algorithm provides quasi-unique solutions, we demonstrate that the mass reconstruction does not depend on our initial guesses.
In Figure~\ref{dif_ini_condition}, we present the A1689 mass reconstruction results when we start the function optimization from four different initial conditions: two flat mass sheets and two NFW profiles.

The 2D mass map comparison shows that the agreement among the four results is excellent in the region constrained by multiple images (to guide eyes, we plot the locations of the multiple images with black dots). The mass map in the outskirt where no SL data exist somewhat depends on the initial conditions, which however is not surprising because no constraints are provided there.

\begin{figure*}
\centering
\begin{center}
    \includegraphics[width=\textwidth]{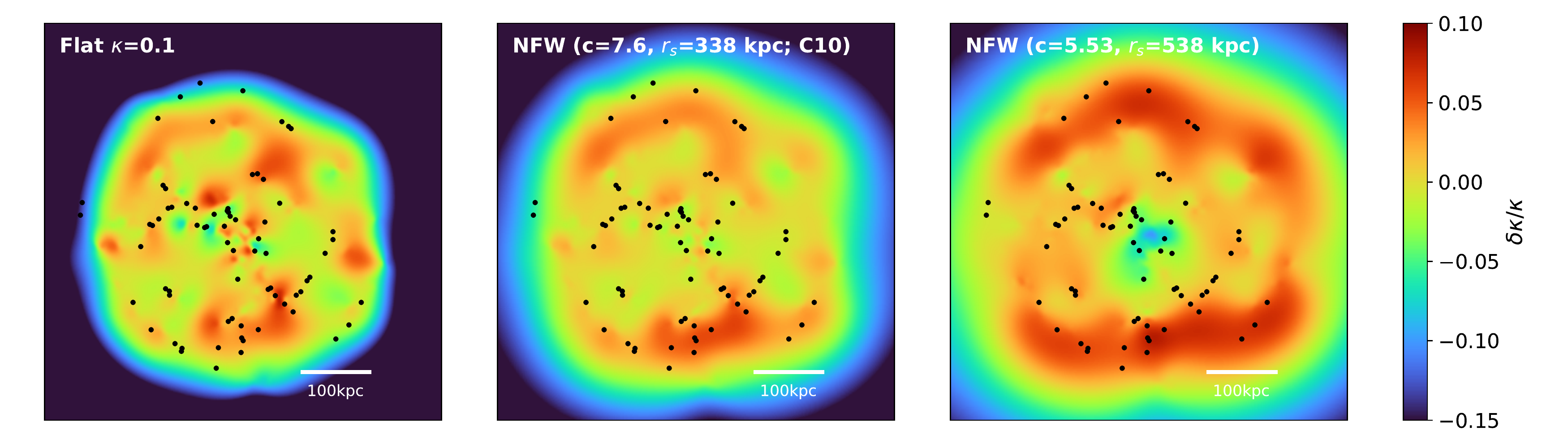}
\end{center}
\caption{Residual mass maps between the three different reconstructions shown in Figure~\ref{dif_ini_condition} and the fiducial one. Left: a residual for the mass map started from a flat mass sheet of $\kappa$=0.1. Middle: a residual for the mass map started from the C10 best-fit NFW result. Right: a residual for the mass map started from our best-fit NFW result. As in Figure~\ref{dif_ini_condition}, we use black dots to mark the locations of the used multiple images. The residual is within $|\Delta \kappa| \lesssim 0.05$ for the region constrained by the multiple images.}
\label{dif_ini_condition_residual}
\end{figure*}

For more quantitative analysis, we compare the radial profiles from these four mass maps in the bottom panel of Figure~\ref{dif_ini_condition}. As indicated by the 2D comparison, no significant difference is found out to $r\sim200$~kpc. Beyond this radius, some small systematic differences are noticeable. In Figure~\ref{dif_ini_condition_residual}, we display the residual mass maps.

%%%%%% comment 32

In summary, we find that {\tt MARS} achieves a quasi-unique solution for the region constrained by the SL data. We note that the optimization requires much longer time if the initial setups are far from the final result (e.g., flat initial conditions).

%%%%%%%%%%%%%%%%%%%%%%%%%%%%%%%%%%%%%%%%%%%%%%%%%%%%%%%%%%%%%%%%%%%%%%%%%%%%%%%%%%%%%%%%%

\subsection{Effect of regularization control parameter} \label{subsec:discussion_r}
\begin{figure*}
\centering
\begin{center}
    \includegraphics[width=0.75\textwidth]{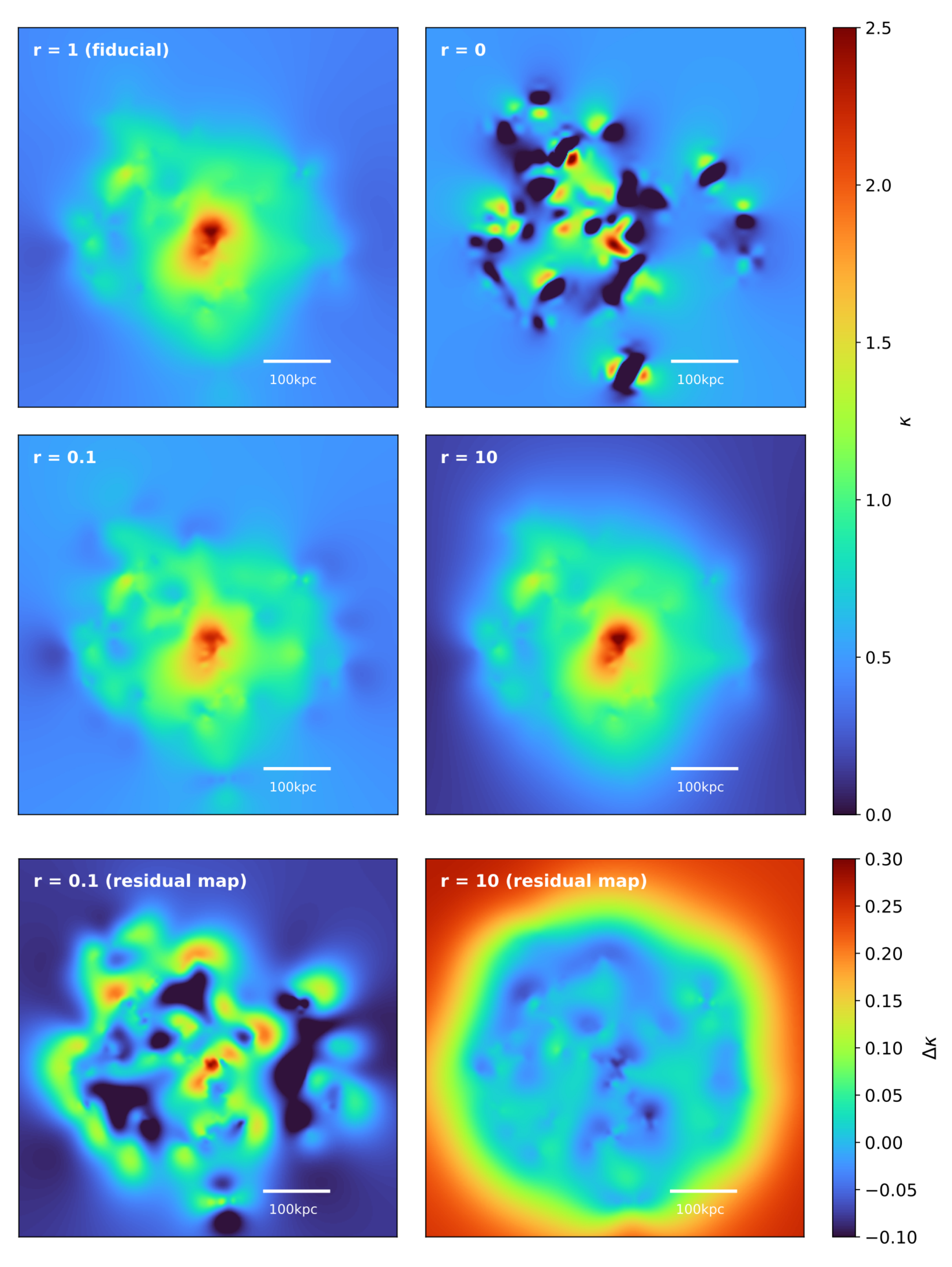}
\end{center}
\caption{Dependence of the reconstruction on the regularization control parameter $r$. We display the A1689 results obtained with different $r$ values. Upper left: result with $r=1$ (fiducial). Upper right: result with $r=0$ (no regularization). Middle left: result with $r=0.1$. Middle right: result with $r=10$. In the bottom panel, we display the residual maps for the $r=0.1$ and $10$ cases. The reference redshift is $z_{f}=9$.}
\label{dif_ini_r}
\end{figure*}

Eqn.~\ref{total_equation} consists of the $\chi^2$ and regularization terms. The regularization term is to prevent overfitting and stabilize the convergence to a quasi-unique solution. In order to illustrate the dependence of the mass map on the regularization control parameter $r$, we display the three ($r=0$, 0.1, 1, and 10) cases in Figure~\ref{dif_ini_r}. 
The $r=1$ case corresponds to our fiducial result, which leads to a mean source plane scatter of $\mytilde10^{-3}\arcsec$. The $r=0.1$ (10) results yield a mean source plane scatter of $\mytilde10^{-4}\arcsec$ ($\mytilde10^{-2}\arcsec$). The differences among the $r=0.1$, 1, and 10 cases are small, but noticeable. As expected, the smaller (larger) $r$ value produces a less (more) smoothed mass map.

The $r=0$ mass map is obtained without any entropy prior and it is not surprising that the resulting mass map shows a number of unphysical, small-scale fluctuations. Because the system is under-determined, the solution is not unique and the result displayed in Figure~\ref{dif_ini_r} is just one of many possible solutions. 

%%%%%%%%%%%%%%%%%%%%%%%%%%%%%%%%%%%%%%%%%%%%%%%%%%%%%%%%%%%%%%%%%%%%%%%%%%%%%%%%%%%%%%%%%

\subsection{Dependence of Error Estimation on Source Position Scatters} \label{subsec:unc_change}
\begin{figure*}
\centering
\includegraphics[width=\textwidth]{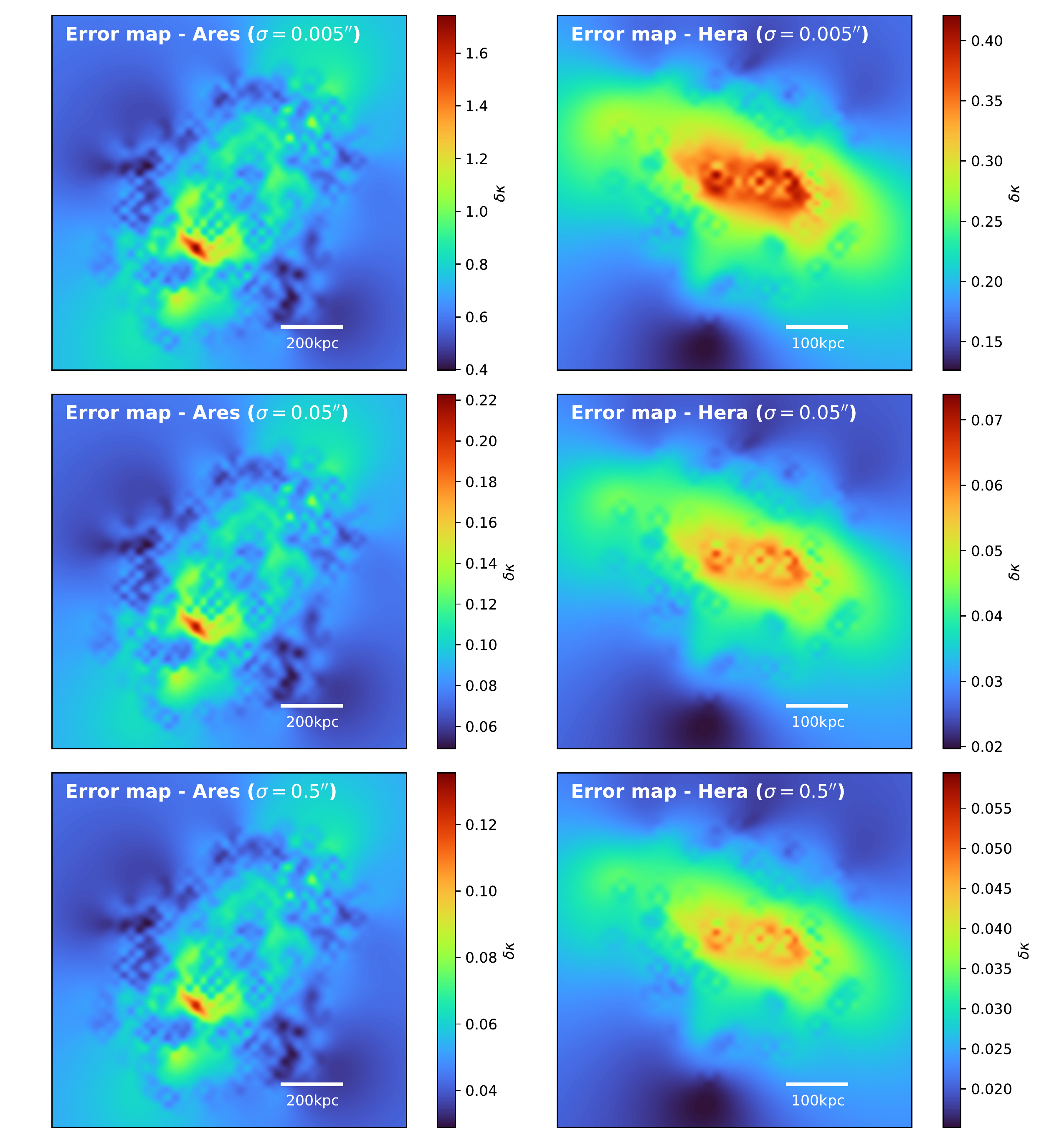}
\caption{Dependence of the uncertainty estimation on the source scatter. Left (right) panels show the uncertainty maps of $Ares$ ($Hera$). $Top$: $\sigma=0.005\arcsec$. $Middle$: $\sigma=0.05\arcsec$. $Bottom$: $\sigma=0.5\arcsec$.}
\label{uncertainty_comparison}
\end{figure*}

\begin{figure*}
\centering
\includegraphics[width=\textwidth]{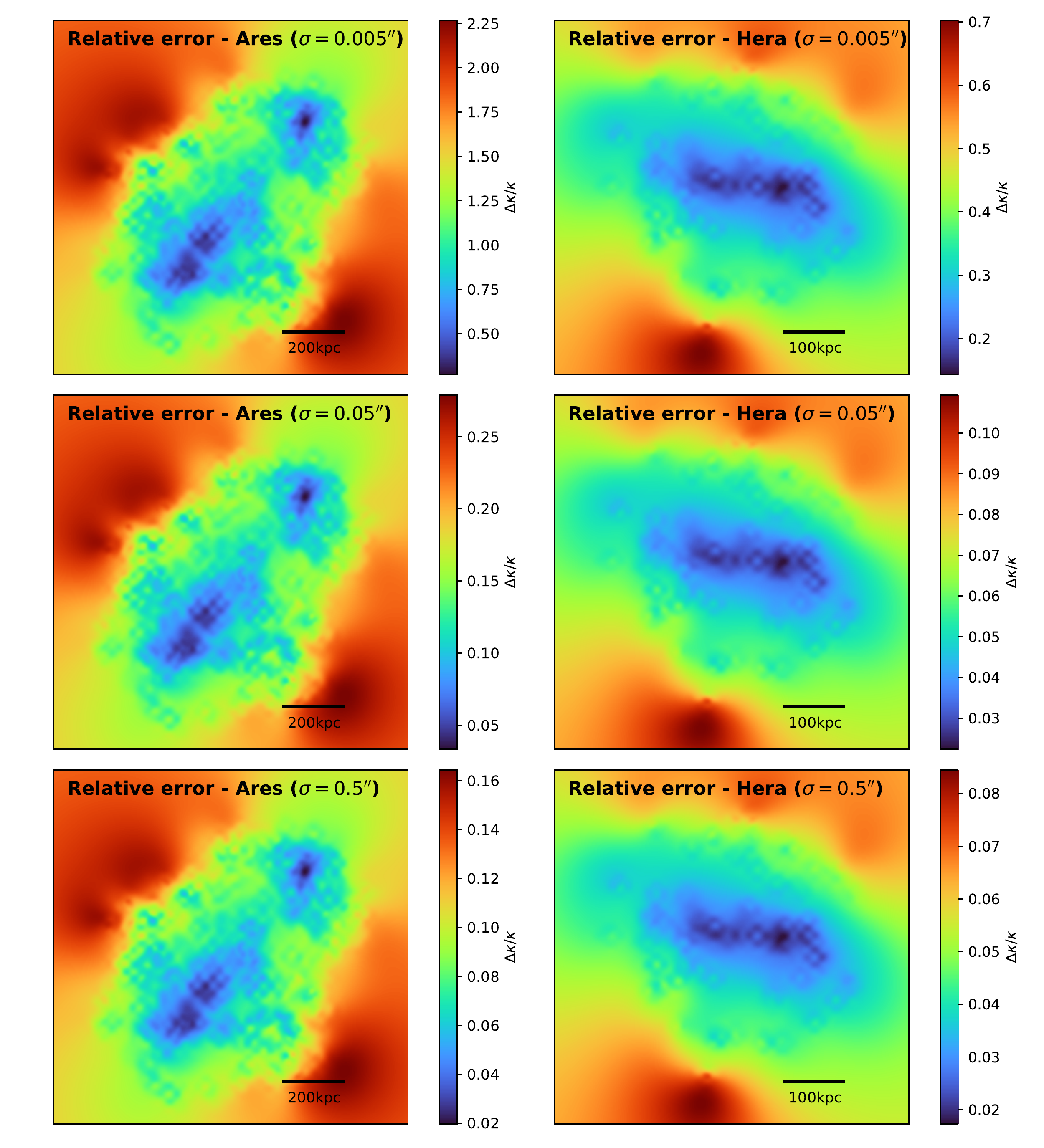}
\caption{Dependence of the relative uncertainty estimation on the source scatter. Left (right) panels show the relative uncertainty maps of $Ares$ ($Hera$). $Top$: $\sigma=0.005\arcsec$. $Middle$: $\sigma=0.05\arcsec$. $Bottom$: $\sigma=0.5\arcsec$.}
\label{relative_uncertainty_comparison}
\end{figure*}

As explained in \textsection\ref{subsec:uncertainties}, we estimate the mass reconstruction errors with the Hessian matrix by assuming that the posteriors follow Gaussian distributions. Another important assumption is the size of the multiple image position scatters in the source plane. For our fiducial result, we assume $\sigma=0.02\arcsec$ after considering the finite resolution of our mass grid. Since it is difficult to define/estimate the exact value for this scatter from the first principle, here we examine the dependence of the mass reconstruction error estimation on the size of the scatter.

In Figure~\ref{uncertainty_comparison}, we show the uncertainty maps of $Hera$ and $Ares$ for the three additional cases: $\sigma=0.005\arcsec$, $0.05\arcsec$, and $0.5\arcsec$. We find that the relative scales across the field are not affected by the assumed scatter values, which is somewhat expected because changing the scatter is equivalent to scaling the likelihood with a simple multiplicative factor.
In terms of the absolute values, we notice significant changes. For $Ares$ the decrease in error is nearly an order of magnitude when we change $\sigma=0.005\arcsec$ to $\sigma=0.05\arcsec$. For the case of $\sigma=0.5\arcsec$, the decrease is about 50\% relative to the $\sigma=0.05\arcsec$ case. The sensitivity is reduced for $Hera$ and we attribute this to the difference in the number of multiple images.

In Figure~\ref{relative_uncertainty_comparison}, we show the relative uncertainty ($\delta \kappa/\kappa$) maps of $Hera$ and $Ares$ for the three cases. Similarly to Figure ~\ref{uncertainty_comparison}, the uncertainty values scale with the size of the source plane scatter.

%%%%%%%%%%%%%%%%%%%%%%%%%%%%%%%%%%%%%%%%%%%%%%%%%%%%%%%%%%%%%%%%%%%%%%%%%%%%%%%%%%%%%%%%%
%%%%%%%%%%%%%%%%%%%%%%%%%%%%%%%%%%%%%%%%%%%%%%%%%%%%%%%%%%%%%%%%%%%%%%%%%%%%%%%%%%%%%%%%%

\section{Conclusion} \label{sec:conclusion}
We have presented a new free-form maximum entropy-regularized SL mass reconstruction method ({\tt MARS}).
The target function that we minimize
consists of two parts:  log-likelihood and regularization terms.
The log-likelihood term is a chi-square function of multiple image positions in the source plane whereas the regularization term is a cross-entropy between the current mass map and the prior. The mass model is represented by a square grid of convergence values and we perform the mass reconstruction in two steps: initial reconstruction with a $50\times50$ ($70\times70$ with the margin) resolution followed by a $100\times100$ ($140\times140$ with the margin) resolution. The method enables us to converge all legitimate multiple image positions in the source plane, suppress spurious small-scale fluctuations due to overfitting, and provide a quasi-unique solution
independently of initial conditions.

We validate our algorithm with 
the publicly available synthetic SL data {\tt FF-SIMS}, which have been used to test a number of existing algorithms. 
We find that the overall performance is on a par with the results from the best-performing LTM methods published in the literature, which is remarkable because the synthetic SL data were generated under the LTM assumption and most non-LTM methods have been shown to perform relatively poorly by a significant margin.
In particular, our mass profile excels in the mass profile reconstruction, agreeing with the truth at the sub-percent level in the region constrained by the SL data; the best-performing LTM methods is known to achieve $\mytilde2$\%-level accuracy whereas most non-LTM methods deviate from the truth by 5-15\%.

For the reconstruction of galaxy-scale substructures, our algorithm underperforms compared to the LTM methods. Also, in the case of $Ares$, the magnification prediction from the the LTM methods is found to be superior to our method. These performance differences might be attributed to the possibility that the LTM approaches can benefit from the use of LTM hypothesis, which was also employed in the synthetic cluster generation.

We apply our method to A1689, which is one of the most studied SL clusters. 
We find that the cluster mass in the SL regime is dominated by the primary halo centered on the BCG and the weaker secondary halo at the bright cluster member $\mytilde160$~kpc northeast of the BCG.
The A1689 radial profile is well-described by an NFW profile with
$c_{200}=5.53\pm0.77$ and $r_s=538^{+90}_{-100}$ kpc, which provides no evidence for overconcentration when only the SL regime is considered.

This  work  is  based  on  observations  created with NASA/ESA Hubble Space Telescope and downloaded from the MiKulski Archive for Space Telescope (MAST) at the Space Telescope Science Institude (STScI).
The current research is supported by the National Research Foundation of Korea under the program 2022R1A2C1003130.

\software{Astropy \citep{astropy2013}, Matplotlib \citep{matplotlib2007}, NumPy \citep{harris2020array}, SciPy \citep{scipy2020}}

%%%%%%%%%%%%%%%%%%%%%%%%%%%%%%%%%%%%%%%%%%%%%%%%%%%%%%%%%%%%%%%%%%%%%%%%%%%%%%%%%%%%%%%%%
%%%%%%%%%%%%%%%%%%%%%%%%%%%%%%%%%%%%%%%%%%%%%%%%%%%%%%%%%%%%%%%%%%%%%%%%%%%%%%%%%%%%%%%%%

\appendix
\section{Multiple Image Catalog of A1689} \label{sec:multiple_images_catalog}
Table~\ref{table1} shows 109 multiple images from 34 source galaxies considered
in this study.
They are classified into the ``gold", ``silver", and ``gold candidate" samples.
The objects in the gold sample have spectroscopic redshifts. 
The silver images have only photometric redshifts and are considered bona fide multiple images
by B05, L07, and C10.
The spectroscopic redshifts of the gold candidate galaxies are unknown to B05, L07, and C10, but are later measured by B16.

\startlongtable
\begin{deluxetable*}{cccccc}
\centering
\tablecaption{Multiple Image Catalog of A1689} \label{table1}
\tablehead {
\colhead{ID} &
\colhead{B05$^{a}$} &
\colhead{RA (J2000)} &
\colhead{DEC (J2000)} &
\colhead{$z$} &
\colhead{class}
}
\renewcommand{\arraystretch}{0.943}
\startdata
1.1 & 1.1 & 13 11 26.452 & -01 19 56.75 & 3.04 & Gold \\
1.2 & 1.2 & 13 11 26.289 & -01 20 00.19 & 3.04 & Gold \\
1.3 & 1.3 & 13 11 29.773 & -01 21 07.43 & 3.04 & Gold \\
1.4 & 1.4 & 13 11 33.066 & -01 20 27.47 & 3.04 & Gold \\
1.5 & 1.5 & 13 11 31.932 & -01 20 05.91 & 3.04 & Gold \\
1.6 & 1.6 & 13 11 29.852 & -01 20 38.50 & 3.04 & Gold \\
2.1 & 2.1 & 13 11 26.524 & -01 19 55.49 & 2.53 & Gold \\
2.2 & 2.2 & 13 11 32.969 & -01 20 25.51 & 2.53 & Gold \\ 
2.3 & 2.3 & 13 11 31.978 & -01 20 07.17 & 2.53 & Gold \\
2.4 & 2.4 & 13 11 29.812 & -01 21 06.05 & 2.53 & Gold \\
2.5 & 2.5 & 13 11 29.881 & -01 20 39.48 & 2.53 & Gold \\
3.1 & 3.1 & 13 11 32.041 & -01 20 27.27 & $5.47^{b}$ & Silver \\
3.2 & 3.2 & 13 11 32.178 & -01 20 33.37 & $5.47^{b}$ & Silver \\
3.3 & 3.3 & 13 11 31.703 & -01 20 55.99 & $5.47^{b}$ & Silver \\
4.1 & 4.1 & 13 11 32.175 & -01 20 57.37 & 1.16 & Gold \\
4.2 & 4.2 & 13 11 30.528 & -01 21 12.02 & 1.16 & Gold \\
4.3 & 4.3 & 13 11 30.758 & -01 20 08.25 & 1.16 & Gold \\
4.4 & 4.4 & 13 11 26.285 & -01 20 35.40 & 1.16 & Gold \\
4.5 & 4.5 & 13 11 29.837 & -01 20 29.45 & 1.16 & Gold \\
5.1 & 5.1 & 13 11 29.064 & -01 20 48.64 & 2.60 & Gold \\
5.2 & 5.2 & 13 11 29.224 & -01 20 44.24 & 2.60 & Gold \\
5.3 & 5.3 & 13 11 34.120 & -01 20 20.96 & 2.60 & Gold \\
6.1 & 6.1 & 13 11 30.755 & -01 19 38.19 & 1.1 & Gold \\
6.2 & 6.2 & 13 11 33.345 & -01 20 12.20 & 1.1 & Gold \\
6.3 & 6.3 & 13 11 32.742 & -01 19 54.49 & 1.1 & Gold \\
6.4 & 6.4 & 13 11 32.478 & -01 19 58.81 & 1.1 & Gold \\
7.1 & 7.1 & 13 11 25.446 & -01 20 51.87 & 4.87 & Gold \\
7.2 & 7.2 & 13 11 30.678 & -01 20 13.99 & 4.87 & Gold \\
7.3 & 7.3 & 13 11 29.824 & -01 20 24.71 & 4.87 & Gold \\
8.1 & 8.1 & 13 11 32.302 & -01 20 51.09 & $2.67^{b}$ & Silver \\
8.2 & 8.2 & 13 11 31.402 & -01 21 05.63 & $2.67^{b}$ & Silver \\
8.3 & 8.3 & 13 11 31.495 & -01 20 14.10 & $2.67^{b}$ & Silver \\
8.4 & 8.4 & 13 11 25.526 & -01 20 20.01 & $2.67^{b}$ & Silver \\
9.1 & 9.1 & 13 11 30.303 & -01 19 48.65 & $5.16^{b}$ & Silver \\
9.2 & 9.2 & 13 11 33.519 & -01 20 50.42 & $5.16^{b}$ & Silver \\
9.3 & 9.3 & 13 11 28.737 & -01 21 15.83 & $5.16^{b}$ & Silver \\
9.4 & 9.4 & 13 11 26.279 & -01 20 26.90 & $5.16^{b}$ & Silver \\
10.1 & 10.1 & 13 11 33.980 & -01 20 51.01 & 1.83 & Gold \\
10.2 & 10.2 & 13 11 28.055 & -01 20 12.61 & 1.83 & Gold \\
10.3 & 10.3 & 13 11 29.316 & -01 20 27.99 & 1.83 & Gold \\
11.1 & 11.1 & 13 11 33.349 & -01 21 06.73 & 2.5 & Gold \\
11.2 & 11.2 & 13 11 29.056 & -01 20 01.31 & 2.5 & Gold \\
11.3 & 11.3 & 13 11 29.498 & -01 20 26.51 & 2.5 & Gold \\
13.1 & 13.1 & 13 11 32.828 & -01 19 24.44 & $1.02^{b}$ & Silver \\
13.2 & 13.2 & 13 11 32.986 & -01 19 25.83 & $1.02^{b}$ & Silver \\
13.3 & 13.3 & 13 11 33.398 & -01 19 31.27 & $1.02^{b}$ & Silver \\
14.1 & 14.1 & 13 11 29.033 & -01 21 41.82 & 3.4 & Gold \\
14.2 & 14.2 & 13 11 29.461 & -01 21 42.73 & 3.4 & Gold \\
15.1 & 15.1 & 13 11 28.082 & -01 20 15.17 & 1.8 & Gold \\
15.2 & 15.2 & 13 11 34.080 & -01 20 51.33 & 1.8 & Gold \\
15.3 & 15.3 & 13 11 29.239 & -01 20 27.62 & 1.8 & Gold \\
16.1 & 16.1 & 13 11 27.990 & -01 20 25.29 & $2.01^{b}$ & Silver \\
16.2 & 16.2 & 13 11 28.905 & -01 20 28.53 & $2.01^{b}$ & Silver \\
16.3 & 16.3 & 13 11 34.400 & -01 20 46.40 & $2.01^{b}$ & Silver \\
17.1 & 17.1 & 13 11 30.662 & -01 20 24.87 & 2.6 & Gold \\
17.2 & 17.2 & 13 11 30.392 & -01 20 27.83 & 2.6 & Gold \\
17.3 & 17.3 & 13 11 24.979 & -01 20 41.84 & 2.6 & Gold \\
18.1 & 18.1 & 13 11 28.245 & -01 20 09.58 & 1.8 & Gold \\
18.2 & 18.2 & 13 11 33.820 & -01 20 54.58 & 1.8 & Gold \\
18.3 & 18.3 & 13 11 29.364 & -01 20 27.39 & 1.8 & Gold \\
19.1 & 19.1 & 13 11 31.634 & -01 20 22.64 & 2.6 & Gold \\
19.2 & 19.2 & 13 11 25.241 & -01 20 20.05 & 2.6 & Gold \\
19.3 & 19.3 & 13 11 31.958 & -01 20 59.38 & 2.6 & Gold \\
19.4 & 19.4 & 13 11 32.040 & -01 20 57.47 & 2.6 & Gold \\
21.1 & 21.1 & 13 11 31.027 & -01 20 45.82 & $1.78^{b}$ & Silver \\
21.2 & 21.2 & 13 11 30.819 & -01 20 44.91 & $1.78^{b}$ & Silver \\
21.3 & 21.3 & 13 11 25.250 & -01 20 11.28 & $1.78^{b}$ & Silver \\
22.1 & 22.1 & 13 11 29.694 & -01 20 08.84 & 1.7 & Gold \\
22.2 & 22.2 & 13 11 29.617 & -01 20 23.76 & 1.7 & Gold \\
22.3 & 22.3 & 13 11 32.420 & -01 21 15.94 & 1.7 & Gold \\
23.1 & 23.1 & 13 11 29.533 & -01 20 10.08 & $2^{b}$ & Silver \\
23.2 & 23.2 & 13 11 29.558 & -01 20 22.96 & $2^{b}$ & Silver \\
23.3 & 23.3 & 13 11 32.662 & -01 21 15.26 & $2^{b}$ & Silver \\
24.1 & 24.1 & 13 11 29.192 & -01 20 56.22 & 2.6 & Gold \\
24.2 & 24.2 & 13 11 32.059 & -01 19 50.56 & 2.6 & Gold \\
24.3 & 24.3 & 13 11 30.290 & -01 19 34.21 & 2.6 & Gold \\
24.4 & 24.4 & 13 11 33.719 & -01 20 19.91 & 2.6 & Gold \\
26.1 & 26.1 & 13 11 25.153 & -01 20 32.74 & 0.96 & Silver \\
26.2 & 26.2 & 13 11 31.326 & -01 20 25.28 & 0.96 & Silver \\
26.3 & 26.3 & 13 11 30.242 & -01 20 32.63 & 0.96 & Silver \\
27.1 & 27.1 & 13 11 25.173 & -01 20 33.16 & $1.1^{b}$ & Silver \\
27.2 & 27.2 & 13 11 31.369 & -01 20 24.64 & $1.1^{b}$ & Silver \\
27.3 & 27.3 & 13 11 30.192 & -01 20 32.94 & $1.1^{b}$ & Silver \\
28.1 & 28.1 & 13 11 28.301 & -01 20 10.86 & $5.45^{b}$ & Silver \\
28.2 & 28.2 & 13 11 34.262 & -01 21 00.01 & $5.45^{b}$ & Silver \\
29.1 & 29.1 & 13 11 29.226 & -01 20 58.06 & 2.5 & Gold \\
29.2 & 29.2 & 13 11 30.022 & -01 19 34.20 & 2.5 & Gold \\
29.3 & 29.3 & 13 11 32.177 & -01 19 52.89 & 2.5 & Gold \\
29.4 & 29.4 & 13 11 33.626 & -01 20 20.64 & 2.5 & Gold \\
30.1 & 30.1 & 13 11 32.420 & -01 19 19.80 & 3.0 & Gold \\
30.2 & 30.2 & 13 11 33.185 & -01 19 26.07 & 3.0 & Gold \\
30.3 & 30.3 & 13 11 33.662 & -01 19 32.73 & 3.0 & Gold \\
32.1 & L07 & 13 11 32.191 & -01 20 03.53 & 3.0 & Gold \\
32.2 & L07 & 13 11 33.216 & -01 20 20.90 & 3.0 & Gold \\
32.3 & L07 & 13 11 29.589 & -01 21 02.83 & 3.0 & Gold \\
32.4 & L07 & 13 11 29.804 & -01 20 43.35 & 3.0 & Gold \\
33.1 & L07 & 13 11 28.449 & -01 21 00.66 & 4.58 & Gold \\
33.2 & L07 & 13 11 34.653 & -01 20 33.60 & 4.58 & Gold \\
35.1 & L07 & 13 11 28.560 & -01 20 59.35 & 1.9 & Gold \\
35.2 & L07 & 13 11 33.953 & -01 20 32.52 & 1.9 & Gold \\
35.3 & L07 & 13 11 29.431 & -01 20 34.66 & 1.9 & Gold \\
36.1 & L07 & 13 11 31.566 & -01 19 45.94 & 3.0 & Gold \\
36.2 & L07 & 13 11 31.686 & -01 19 47.39 & 3.0 & Gold \\
40.1 & L07 & 13 11 30.260 & -01 20 12.04 & 2.52 & Gold \\
40.2 & L07 & 13 11 26.176 & -01 21 03.29 & 2.52 & Gold \\
46.1 & C10 & 13 11 31.669 & -01 20 47.19 & $3.48^{c}$ & Gold Candidate \\
46.2 & C10 & 13 11 24.959 & -01 20 13.98 & $3.48^{c}$ & Gold Candidate \\
50.1 & C10 & 13 11 32.580 & -01 20 43.60 & $4.27^{c}$ & Gold Candidate \\
50.2 & C10 & 13 11 31.021 & -01 21 09.08 & $4.27^{c}$ & Gold Candidate \\
50.3 & C10 & 13 11 31.660 & -01 20 13.66 & $4.27^{c}$ & Gold Candidate \\
\enddata
\tablenotetext{a}{We follow the numbering scheme of B05.}
\tablenotetext{b}{Photometric redshifts.}
\tablenotetext{c}{Spectroscopic redshift data from B16.}
\end{deluxetable*}

%%%%%%%%%%%%%%%%%%%%%%%%%%%%%%%%%%%%%%%%%%%%%%%%%%%%%%%%%%%%%%%%%%%%%%%%%%%%%%%%%%%%%%%%%

\section{Mass Reconstruction of A1689 with Additional Multiple Images} \label{sec:result_a1689_etc}
\begin{figure*}
\centering
\includegraphics[width=0.72\textwidth]{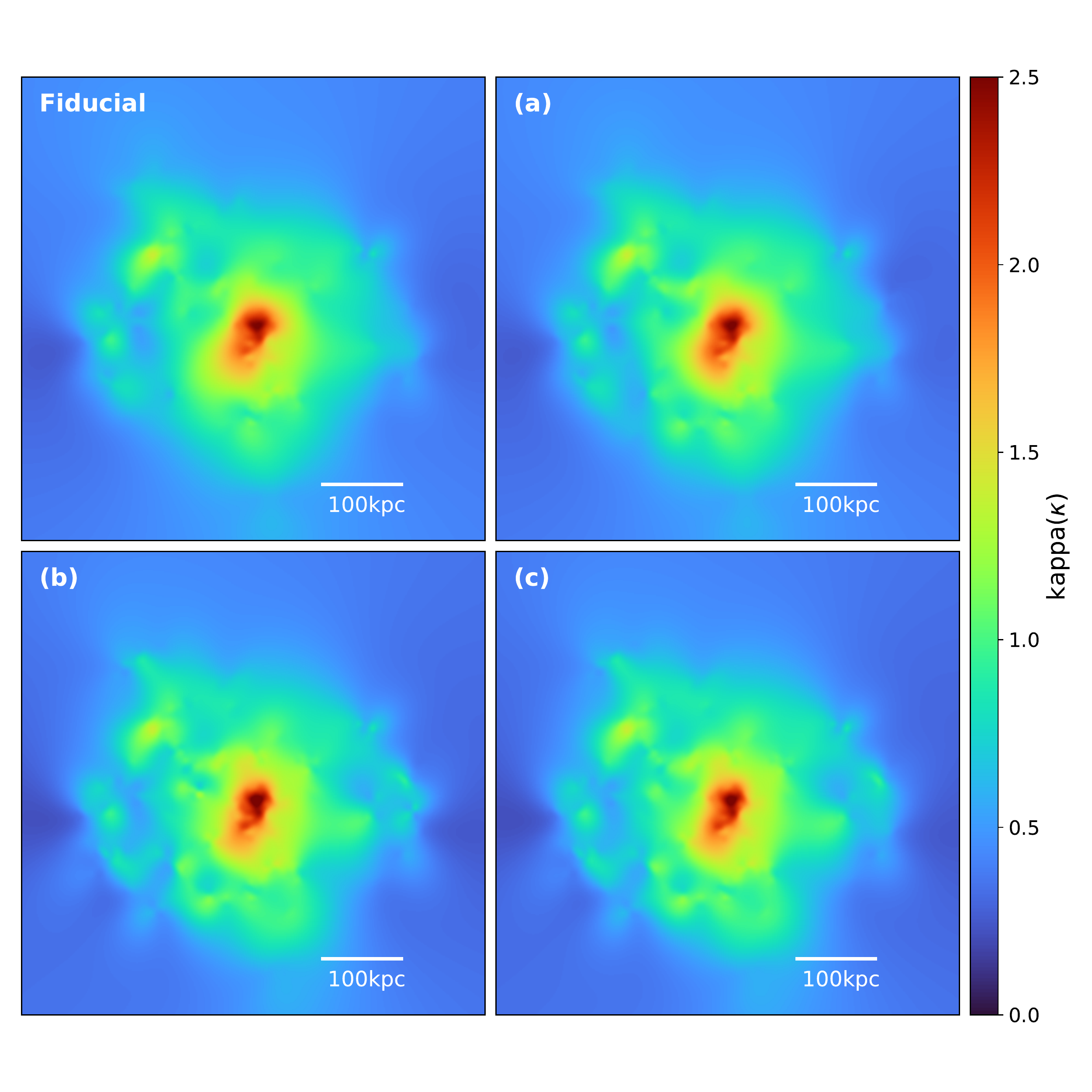}
\caption{Mass reconstruction of A1689 with different source selections. $Fiducial$: reconstruction with the ``gold" sample. $a$: reconstruction with the ``gold" + ``gold candidate" samples. 
$b$:  reconstruction with the ``gold" + ``gold candidate" + ``silver" samples. $c$: same as $b$ except source IDs 26 and 27 are removed. The reference redshift is $z_{f}=9$.}
\label{A1689_additional_images}
\end{figure*}

\begin{figure*}
\centering
\includegraphics[width=\textwidth]{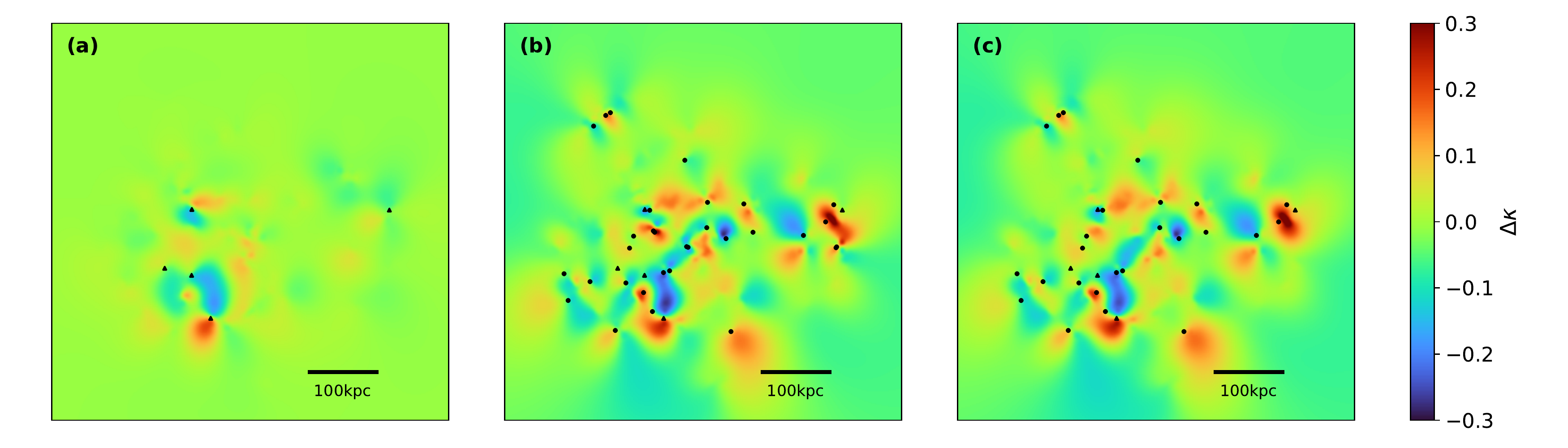}
\caption{Residual mass maps between the three different reconstructions in Figure~\ref{A1689_additional_images} and the fiducial one. Left: a residual for the mass map with the ``gold" + ``gold candidate" samples. Middle: a residual for the mass map with the ``gold" + ``gold candidate" + ``silver" samples. Right: a residual for the mass map with same as middle one except source IDs 26 and 27 are removed. Markers indicate the position of additional multiple images. Black triangles (dots) show the gold candidate (silver) multiple images.}
\label{A1689_additional_images_residual}
\end{figure*}

The fiducial result for A1689 is obtained with the ``gold" samples (\textsection\ref{subsec:real_cluster}). Here we investigate how the result changes when we consider ``silver" and ``gold candidate" samples.
With the selection of the ``gold" + ``gold candidate" samples, we are able to successfully delens all
multiple images to converge in the source plane. The resulting mass map (Figure~\ref{A1689_additional_images}a) is also very similar to our fiducial case.
When we use all three samples (``gold" + ``gold candidate" + ``silver''), we find that 
the two sources (IDs 26 and 27) in the ``silver" sample cannot be delensed to converge in the source plane.
We suggest three possibilities, which are not mutually exclusive.  
First, the two multiple systems are very close to each other on the image plane (see Figure~\ref{fig:A1689_all}). Thus, the resolution of our grid might not be sufficiently large to distinguish the two systems.
Second, we might have used inaccurate redshifts for these images. As we mentioned in \textsection\ref{subsec:real_cluster}, the "silver" multiple images have only photometric redshifts.
Thus, it is a legitimate concern that for some sources
their photo-$z$ redshift might be catastrophically different from the true values \citep{2014ApJS..214...24S}.
Third, these multiple images may not be real multiple images. \citet{2015MNRAS.446..683D} state that the authors removed sources 26 and 27 from their catalog because the images show different colors in their newly added IR data. 
Nevertheless, we find that the resulting mass reconstruction (Figure~\ref{A1689_additional_images}b) is still similar to our fiducial version, although some additional small-scale ``wrinkles" are introduced.
Finally, we carry out mass reconstruction using the ``gold" + ``gold candidate" + ``silver''
without these two problematic sources (Figure~\ref{A1689_additional_images}c) and find that all multiple images are successfully delensed within $~\sim 0.001 \arcsec$ in the source plane. Again, as expected, the resulting mass map is in good agreement with the previous cases.

In Figure ~\ref{A1689_additional_images_residual}, we show the residual maps with respect to the fiducial result. The positions of multiple images also are marked. We can clearly see that the region where multiple images located shows larger differences from the fiducial one.

%%%%%%%%%%%%%%%%%%%%%%%%%%%%%%%%%%%%%%%%%%%%%%%%%%%%%%%%%%%%%%%%%%%%%%%%%%%%%%%%%%%%%%%%%

\section{Source Image Reconstruction with the A1689 mass model}\label{source_reconstruction_a1689}
\begin{figure*}
\centering
\includegraphics[width=0.9\textwidth]{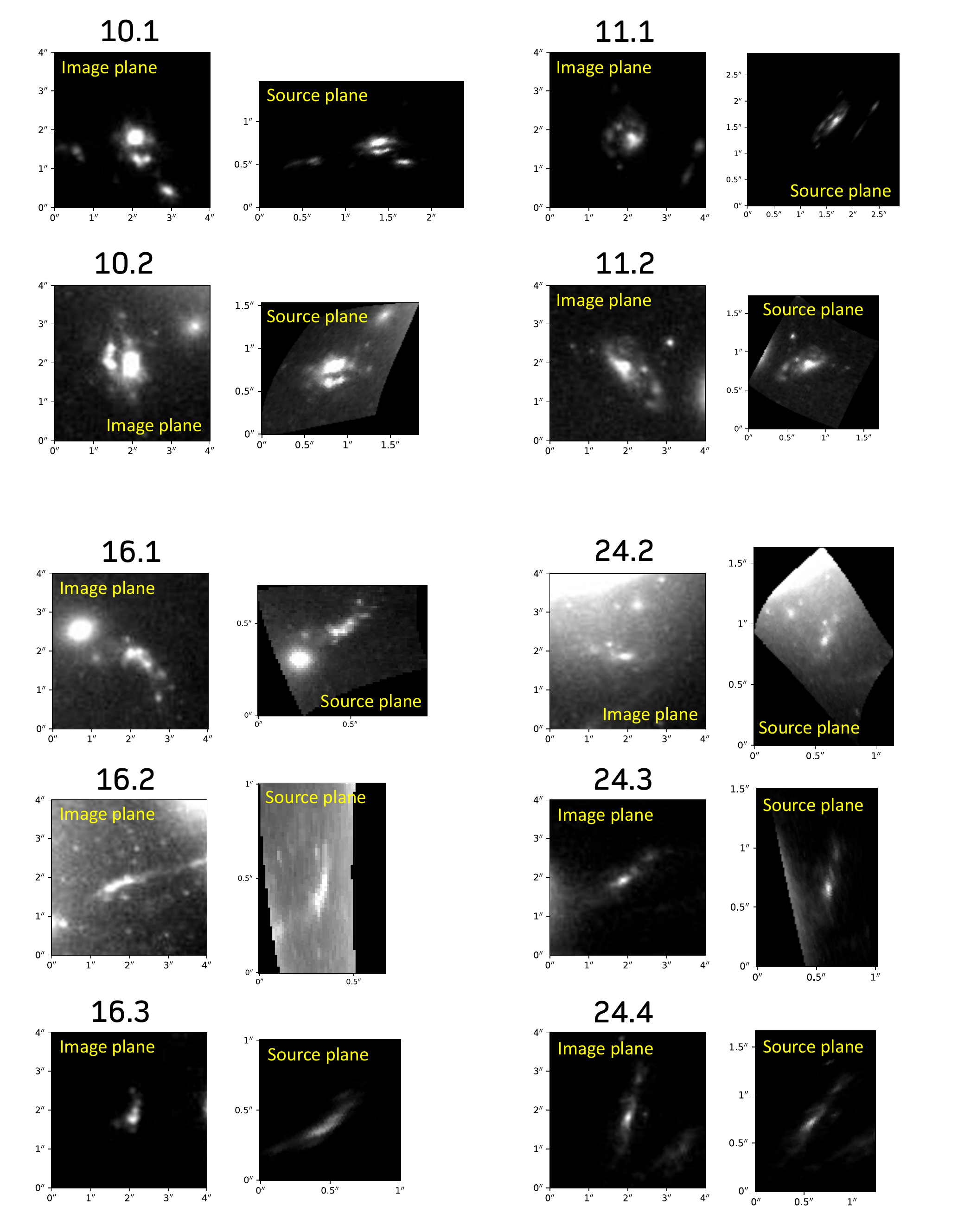}
\caption{Source plane image reconstruction. We displays the results for the four sources: IDs 10, 11, 16, and 24, which have a distinctive morphology in the deep HST/ACS F814W image.
For each ID, the image on the left-hand (right-hand) side is the observed (reconstructed) image. North is up and east is left.}
\label{source_recon}
\end{figure*}

One of the useful applications of the SL mass reconstruction results is the source plane image reconstruction, which not only provides the model verification, but also enables us to use strong lensing clusters as a cosmic lens for observations of more distant universe.
In this appendix, we present source image reconstruction of the multiple images in A1689.

Figure~\ref{source_recon} displays the reconstructed images of the four sources: IDs 10, 11, 16, and 24.
Although in principle, we can apply this delensing to any multiple image system, these four sources have a distinctive morphology in the deep F814W image. 
Among our mass models, we choose the deflection angle obtained from the ``gold" + "gold candidate" + "silver" samples (i.e., the model shown in Figure~\ref{A1689_additional_images}b). 
We find that overall the reconstructed images from the same sources show reasonable agreements in morphological features, orientations, parities, and sizes, which is remarkable because our reconstruction  is only based on the source center position convergence.

%%%%%%%%%%%%%%%%%%%%%%%%%%%%%%%%%%%%%%%%%%%%%%%%%%%%%%%%%%%%%%%%%%%%%%%%%%%%%%%%%%%%%%%%%
\section{Giant arc reproduction with the A1689 mass model}\label{arc_reproduce}

\begin{figure*}
\centering
\includegraphics[width=0.85\textwidth]{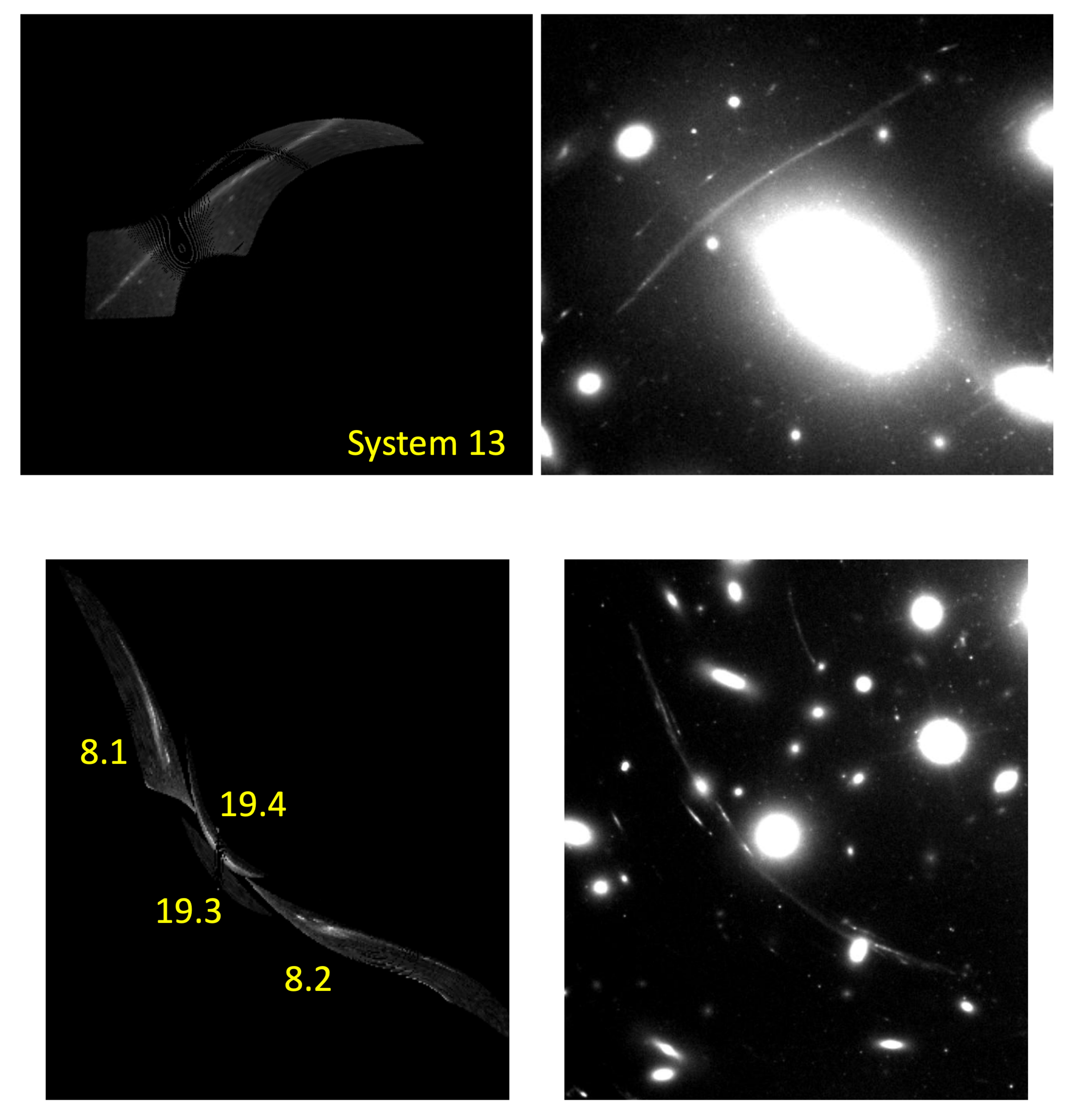}
\caption{Giant arc reproduction in the A1689 field. We choose the two giant arcs that have a distinct morphology in the deep HST/ACS F814W image. The images on the left-hand (right-hand) side are the reproduced (observed) arcs. Although not perfect, the resemblance is remarkable, especially given that {\tt MARS} only uses the source plane scatter as  constraints without utilizing the image plane morphology. }
\label{fig:arc_reproduce}
\end{figure*}

In principle, a perfect lens model should be able to reproduce all observed SL features. However, in general, it is difficult to predict the morphology of the giant arcs precisely because their positions are  close to the critical curves and thus the result is highly sensitive to small perturbations. Since MARS is a grid-based free-form algorithm, it is not best poised to perform arc reproduction because of its limited resolution. Nevertheless, it would be interesting to investigate how well MARS can reproduce the giant arcs in the A1689 field.

In Figure~\ref{fig:arc_reproduce}, we display our reconstruction of the two giant arcs, which have a distinct morphology in the deep F814W image.
The first arc is from system 13 whereas the second arc consists of systems 8 and 19. The overall extent and morphology of the two arcs resemble the observed ones. Given that MARS cannot include galaxy-scale substructures and the two arcs are positioned near the bright galaxies, we believe that the resemblance, although it is not perfect, is rather remarkable. We stress that MARS only minimizes the source plane scatter without using the image plane morphology of the multiple images.

\bibliographystyle{apj}
\bibliography{mypaper}

\end{document}